\documentclass[12pt,aps,prx,eqsecnum,onecolumn,notitlepage,%
nofootinbib,preprintnumbers,superscriptaddress,longbibliography,%
amsmath,amssymb]{revtex4-1}
\usepackage{bm}
\usepackage{graphicx}
\usepackage{color}
\usepackage[colorlinks=true,pdfstartview=FitV,citecolor=blue,urlcolor=blue]{hyperref}

\newcommand{\be}{\begin{equation}}
\newcommand{\ee}{\end{equation}}
\newcommand{\bear}{\begin{eqnarray}}
\newcommand{\eear}{\end{eqnarray}}
\newcommand{\ba}{\begin{array}}
\newcommand{\ea}{\end{array}}

\newcommand{\calL}{\mathcal{L}}
\newcommand{\bj}{\boldsymbol{j}}
\newcommand{\bx}{\boldsymbol{x}}
\newcommand{\bA}{\boldsymbol{A}}
\newcommand{\bB}{\boldsymbol{B}}
\newcommand{\bD}{\boldsymbol{D}}
\newcommand{\bE}{\boldsymbol{E}}
\newcommand{\bL}{\boldsymbol{L}}
\newcommand{\bP}{\boldsymbol{P}}
\newcommand{\bnabla}{\boldsymbol{\nabla}}
\newcommand{\tilmu}{\tilde{\mu}}
\newcommand{\tillam}{\tilde{\lambda}}

\newcommand{\tilq}{\tilde{q}}
\newcommand{\muB}{\mu_{\text{B}}}

\begin{document}
\title{Classification of Magnetic Vortices\\
  by Angular Momentum Conservation}

\preprint{KEK-TH-2268, J-PARC-TH-0231}

\author{Kenji Fukushima}  \email{fuku@nt.phys.s.u-tokyo.ac.jp}
\affiliation{Department of Physics, The University of Tokyo,
  7-3-1 Hongo, Bunkyo-ku, Tokyo 113-0033, Japan}

\author{Yoshimasa Hidaka}  \email{hidaka@post.kek.jp}
\affiliation{Institute of Particle and Nuclear Studies, KEK, 1-1 Oho,
  Tsukuba, Ibaraki 305-0801 Japan}
\affiliation{RIKEN iTHEMS, RIKEN, 2-1 Hirosawa, Wako,
  Saitama 351-0198, Japan}
          
\author{Ho-Ung Yee}  \email{hyee@uic.edu}
\affiliation{Department of Physics, University of Illinois, Chicago,
  Illinois 60607, U.S.A.}

\begin{abstract}
  Superfluid vortices are quantum excitations carrying quantized
  amount of orbital angular momentum in a phase where global symmetry
  is spontaneously broken.  We address a question of whether magnetic
  vortices in superconductors with dynamical gauge fields can carry
  nonzero orbital angular momentum or not.  We discuss the angular
  momentum conservation in several distinct classes of examples from
  crossdisciplinary fields of physics across condensed matter, dense
  nuclear systems, and cosmology.  The angular momentum carried by
  gauge field configurations around the magnetic vortex plays a
  crucial role in satisfying the principle of the conservation law.
  Based on various ways how the angular momentum conservation is
  realized, we provide a general scheme of classifying magnetic
  vortices in different phases of matter.
\end{abstract}
\maketitle

\section{Introduction}

When a superfluid where a macroscopic condensate of identical bosons
is formed is under rotation, superfluid vortices emerge and each
microscopic constituent carries the same amount of orbital angular
momentum, i.e., an integer multiple of fundamental quanta, $\hbar$.
This is a remarkable way to store macroscopic amount of angular
momentum in a highly coherent quantum state.  Not only in table-top
physical systems of superfluids such as $^4$He, the superfluid vortex
can also be an important constituent in rotating nuclear matter found
inside neutron star, where the extremely high matter density causes a
nonzero order parameter that signifies spontaneous breaking of
global baryon number $U(1)_{\rm B}$ symmetry.  In a more interesting
scenario of dense quark matter in quantum chromodynamics (QCD), this
order parameter is also responsible for the superconducting phase of
color gauge interactions, most likely the color-flavor-locked (CFL)
superconducting phase~\cite{Alford:1998mk}.  There exist highly
nontrivial vortices in the CFL phase, called non-Abelian CFL
vortices~\cite{PhysRevD.73.074009}, that involve both dynamics of the
global baryon and the local color gauge symmetries; see
Ref.~\cite{Eto:2013hoa} for a comprehensive review.

As seen in many interesting examples including vortices in CFL quark
matter, some of which we will study in later discussions, the
symmetries involved in vortex contents are entirely or partially gauge
symmetries.  The prototypical example is of course the magnetic vortex
in Type-II superconductors.  In these cases the vortex profile is
fundamentally different from that of the purely superfluid vortex;  a
magnetic flux is threaded into the vortex core.  Among many
differences between a superfluid vortex and a gauged magnetic vortex,
one may specifically ask about the angular momentum they carry.
Surprisingly to us, we find that this simple question has not been
properly addressed in the literature.  
As we try to answer the question in
various examples across different fields of physics, we discover
surprisingly diverse situations.  It is the purpose of this article to
present a compelling list of examples where the answers are quite
different from each other, and also to provide an overarching physics
explanation of why the answers can be so diverse.  We will demonstrate
that the angular momentum conservation offers a key guiding principle
to understand the physics origin of the different answers.  Our
detailed analysis in the main text shows that the angular momentum
carried by not only the matter sector of the system but also the
dynamical gauge fields surrounding the vortex should be considered in
order to fulfill the principle of angular momentum conservation.
Building upon this principle, we attempt a general classification
scheme of magnetic vortices in different phases of matter, that can
hopefully be applied to other physical systems.

A natural starting point of our discussion lies in the vortices in
Type-II superconductors.  Quite generally, it is easy to see that the
vortex should carry a nonvanishing orbital angular momentum.  Due to
one of the Maxwell equations, $\bnabla\times\bB=\bj$ (where we chose a
natural unit in which the magnetic constant $\mu_0$ is the unity), a
smooth and finite ranged profile of magnetic flux means the existence
of azimuthal component of the current density $\bj$.  Under a fairly
general assumption that the charge and momentum carriers are
nonrelativistic quasi-electrons in the conduction band forming the
Fermi surface with superconducting gap, the current and the momentum
are linearly related as an operator relation, holding for all states:
$\bj = -{e\over m}\bP$, where $\bP$ is the momentum density operator,
$m$ is the effective mass of conduction electrons, and $-e$ is the
charge of electrons.  Since $\bj\neq \bm{0}$, we have $\bP\neq \bm{0}$ and the
finite sized vortex should carry a finite angular momentum by
$\bL = \int_x \bx\times \bP\neq \bm{0}$.

The linear relation between the current and the momentum for
nonrelativistic electrons is a consequence of Galilean invariance, and
is not necessarily universal.  Even though the dispersion relation
deviates from the nonrelativistic Galilean invariant one, the current
and the momentum are still negatively correlated, and there is no
reason to exclude nonvanishing angular momentum.  This discussion also
implies that the quantization of the angular momentum in units of
$\hbar$ may not be universal.  To complicate the situation more
nontrivially, some vortices may also carry a localized electric
charge~\cite{PhysRevLett.75.1384,CLAYHOLD2003272}, and the resulting
electromagnetic (EM) field around such a vortex gives rise to a
nonvanishing Poynting vector around the vortex core axis.  The total
angular momentum should then include a contribution from the
EM field around the vortex.

Although the above features are robust, one can consider the following
thought-experiment, that is somewhat similar to Feynman's angular
momentum paradox.  One places a solenoid below a superconductor
sample, and turns on an external magnetic field to create magnetic
vortices piercing the superconductor.  The process can be implemented
in azimuthal symmetric way, and should not change the total angular
momentum which is zero initially.  Since the created vortices carry
finite amount of angular momentum, where can the compensating angular
momentum be found?

The answer to this question is easy to guess: the background of solid
crystal and the electrons in filled valence bands should carry the
compensating angular momentum.  Their inertia is infinitely large and
their rotation may not be detectable, but the torque acting on them
during the vortex creation process should impart to them precisely the
negative amount of the angular momentum of the created vortices.  In
the next section we will be able to confirm this quantitatively in a
concrete model which is simple and yet general enough to carry out the
analysis of charged magnetic vortices.  In this case the angular
momentum carried by EM field also needs to be counted in the total
angular momentum, and the angular momentum conservation holds true
quite nontrivially only after including this EM contribution.  We note
that the EM field is localized around and attached to the vortex, so
one should think of it as a part of the vortex profile.

An obvious next question as a continuation of the above
thought-experiment of creating the vortex by a hypothetical solenoid is:
what would happen in a system that has no background matter to absorb
the angular momentum?  A concrete example of such system is provided
by a relativistic field theory which is self-consistent by itself
without any other degrees of freedom:  it could be identified as the
electroweak sector of the Standard Model with Higgs field condensate,
or more simply, a theoretical model by Nielsen and
Olesen~\cite{Nielsen:1973cs}.  For this class of examples, our
previous argument of angular momentum conservation becomes powerful
enough to dictate that any magnetic vortices, either charged or not,
should carry zero angular momentum.  We call them
``spinless vortices.''  We will show that this statement is indeed
true for the Nielsen-Olesen model.  In showing this for the charged
vortex case, it is again critical to include the EM or gauge
contribution to the total angular momentum.  For a similar conclusion
for the dyonic solitons, see
Refs.~\cite{VanderBij:2001nm,vanderBij:2002sq,Navarro-Lerida:2014bja}.  
We make a remark that Appendix C of a renowned paper,
Ref.~\cite{Julia:1975ff}, contains an erroneous statement on this for
the charged vortex case, which we will rectify.  With this example we
showcase the nontriviality of our argument of angular momentum
conservation.

What would happen if a vortex consists of a combination of magnetic
vortex of gauge symmetry and superfluid vortex of global symmetry?  To
answer this question, we take an example of the
``non-Abelian vortex''~\cite{PhysRevD.73.074009,PhysRevD.78.045002} in
the CFL superconducting phase of dense quark matter, which may be
relevant for the physics of neutron star cores.  The non-Abelian CFL
vortices also play a role in the idea of quark-hadron
continuity~\cite{Schafer:1998ef} in the high density region of QCD
phase diagram~\cite{Hatsuda:2006ps,PhysRevD.86.121704,Alford:2018mqj,Chatterjee:2018nxe}
(see, however, Refs.~\cite{Cherman:2018jir,Hirono:2018fjr,Cherman:2020hbe} for recent
debates on the idea of quark-hadron continuity).
In this example, color symmetry is obviously (non-Abelian) gauged, and
the global symmetry is associated with the $U(1)_{\rm B}$ baryon
number.  For such an object of composite nature, one can imagine a
creation process by an external magnetic field for the gauge symmetry
together with a physical rotation of the superfluid for the global
symmetry.  Our angular momentum conservation argument then predicts
that the total angular momentum should be given only by the superfluid
part of global symmetry without any contribution from the gauged
symmetry part.  We will explicitly confirm this expectation in a
highly nontrivial manner.

There is a logical exception to the above argument for spinless
vortices in a system with no background.  During the above considered
creation process by an external solenoid, a finite amount of angular
momentum may diffuse away to spatial infinity, resulting in an
opposite amount of angular momentum localized around the vortex.  The
total angular momentum is conserved and zero, but the part at infinity
is not visible, and should not be thought of as a contribution to the
angular momentum of the localized vortex.  The vortex then carries a
left-over angular momentum that is finite.  What distinguishes this
case from our first case with background matter is that the opposite
angular momentum to the one carried by the vortex strictly resides at
spatial infinity, or more precisely at the boundary of the system far
away from the vortex.  This makes a contrast to the previous case with
background matter, where the bulk of the background absorbs a finite
angular momentum.

An instructive example of this class of vortex is provided by the
magnetic vortex on the surface of a time-reversal invariant (i.e.,
$T$-invariant) strong topological insulator (TI) in a setup recently
studied in Ref.~\cite{PhysRevLett.121.227001}.  Although the authors
of Ref.~\cite{PhysRevLett.121.227001} considered an interface between
TI and a superconductor, we will focus on the TI part to account
clearly for the physics origin of the net fractional (in units of
$\hbar$) angular momentum of the vortex.  We will show that the total
angular momentum solely arises from the gauge field configuration
surrounding the vortex on the TI surface, without any TI matter
contributions.  We will argue for this peculiar feature that the
topological nature of the TI is responsible for moving apart a finite
angular momentum to the (infinitely separated) boundary, which
characterizes this class of example.  Ubiquitous topological vortices
with fractional angular momentum in topological phases of matter as
found in Refs.~\cite{Seiberg:2016rsg,Seiberg:2016gmd,Senthil:2018cru}
should belong to this class of magnetic vortex.

The final class of vortex in our classification is delineated by the
last logical possibility:  a vortex may \textit{not} be created by our
thought-experiment with external solenoid in a way that conserves
angular momentum, and additional operations to violate angular
momentum conservation must be performed to create a vortex.  This
class of vortex is rather exotic and rarely found in the literature:
one example we address in this paper is an object called the ``charged
semilocal vortex'' of Abraham~\cite{Abraham:1992hv}.  Since this class
of vortex simply falls outside of our principle of angular momentum
conservation, they may or may not carry an angular momentum: in our
example the charged semilocal vortex carries a finite angular
momentum.  It is an inhomogeneous profile along the vortex axis that
makes it impossible to create this kind of vortices by simply piercing
an external magnetic flux:  an additional ``twisting'' or ``spinning''
along the axis is needed to create such a vortex profile.

In summary, we have the following distinct classes of vortices in
regard to angular momentum conservation and their creation processes:
\begin{itemize}
\item {\bf Class Ia (spinful vortices)}:  They carry a finite angular
  momentum due to the existence of background matter that can absorb
  angular momentum.  Examples are the vortices in Type-II
  superconductors.

\item {\bf Class Ib (topological vortices)}:  They carry a finite
  angular momentum, but no background matter exists in the bulk.  The
  angular momentum resides only on the boundary.  The angular momentum
  carried by the surrounding gauge fields must be counted for the
  total angular momentum.  Examples are the vortices on the surface of
  topological phases of matter.

\item {\bf Class II (spinless vortices)}:  They do not carry a net
  angular momentum due to the angular momentum conservation.  The
  angular momentum carried by the surrounding gauge fields should be
  included.  Examples are the vortices in relativistic field theories
  and cosmology.

\item {\bf Class III (exotic vortices)}:  They have an inhomogeneous
  profile along the vortex axis, so that they cannot be created by a
  simple procedure of piercing magnetic flux.  They may or may not
  carry angular momentum.  An example is the charged semilocal
  vortex.
\end{itemize}

In the following sections, we present detailed analysis on concrete
examples that belong to each of the above classes, in order.

\section{Class I: Case study of magnetic vortices with nonzero angular
  momentum}

In this section we discuss the case of magnetic vortices that carry a
nonzero angular momentum.  Because the angular momentum should be
conserved as long as rotational symmetry is preserved, the angular
momentum of magnetic vortices, if it is nonzero, should be balanced
with other contributions.  According to the types of such balancing
contributions, we further classify them into two distinct subclasses;
namely, Class~Ia and Class~Ib.

\subsection{Class Ia -- Incomplete cancellation due to background matter}
\label{sec:ClassIa}

The most familiar magnetic vortices in Type-II superconductor
belong to Class~Ia.  The magnetic vortices can carry a nonzero
angular momentum but its value is not quantized in units of $\hbar$,
unlike the angular momentum carried by superfluid vortices.  Explicit
calculations as shown below make clear where the difference appears.

For an explicit demonstration we shall consider a relativistic scalar
field theory in the Higgs phase of U(1) symmetry, so that gauged
magnetic vortices emerge.  We then take the nonrelativistic reduction
and find the equations of motion that are familiar in condensed matter
physics describing magnetic vortices in Type-II superconductors.  The
Lagrangian density we study in the natural unit system ($\hbar=c=1$) reads as
\be
  \calL = (D_\mu \phi)^\dag (D^\mu \phi) - U(\phi)
  +\frac{1}{2}\bE^2- \frac{1}{2}\bB^2
  - q A_0\,,
  \label{eq:Lscalar}
\ee
where $D_\mu=\partial_\mu+ie n_e A_\mu$ is the covariant derivative,
with $n_e$ being the electric charge carried by $\phi$ in 
units of $e>0$.  As usual, $\bE=-\bnabla A_0 -\partial_0 \bA$ and
$\bB=\bnabla\times \bA$ are electric and magnetic fields,
respectively.  We should choose $n_e=-2$ for the Cooper pair in
electron superconductivity.  The last term, $q A_0$, with a background
charge density $q(x)$, is introduced to keep the total electric charge
neutrality, which we will simply refer to as the ``background'' in the
following.  For example, in a solid with conduction electrons,
positively charged ions in the crystal and other electrons in valence bands neutralize the whole system.
We also note that a finite chemical potential $\mu$ will be introduced
by replacing $ie n_e A_0 \to ie n_e A_0 - i\mu$.  The potential,
$U(\phi)$, is chosen to have a nonzero condensate of $\phi$ in the
Higgs phase, the simplest choice of which would be a polynomial form:
\be
U(\phi) = -\lambda_2 |\phi|^2 + \frac{\lambda_4}{2} (|\phi|^2)^2\,.
\ee
The equations of motion from the Lagrangian are given by
\begin{align}
  & -(D_\mu D^\mu)\phi + \lambda_2 \phi - \lambda_4 |\phi|^2\, \phi = 0\,,
  \label{eq:eom_phi}\\
  & -\partial_0\bE + \bnabla\times \bB
     + ie n_e \bigl[ (\bD\phi)^\dag \phi - \phi^\dag (\bD\phi) \bigr]=0\,,   \label{eq:eom_A}\\
  & \bnabla\cdot  \bE
     + ie n_e \bigl[ (D_0\phi)^\dag \phi - \phi^\dag D_0 \phi \bigr] - q=0\,,
    \label{eq:eom_A0}
\end{align}
with $(\bD)^i\equiv D^i$, where we note that
$\partial^i=\partial/\partial x_i=-\partial/\partial x^i$ in our
metric convention $(+,-,-,-)$.  The magnetic vortices we consider are
the static solutions of the above equations of motion, so we drop the
time derivative terms in the below.  Then, Eq.~\eqref{eq:eom_phi}
takes a form of
\be
  \bigl[ (\mu - e n_e A^0)^2 + \bD^2 + \lambda_2\bigr]\phi
  -\lambda_4 |\phi|^2 \phi = 0\,.
\label{eq:eom_phi2}
\ee
 Instead
of solving this equation directly, we would like to make the problem
close to a more conventional situation in condensed matter physics, by
considering the nonrelativistic reduction.  Because the
nonrelativistic energy is measured from the rest mass energy $m$,
we should split the
mass term and rescale the field as
\be 
  \mu \to m+\tilmu\,,\qquad 
  \lambda_2 \to -m^2 + 2m\tillam_2\,,\qquad 
  \phi \to \frac{\psi}{\sqrt{2m}}\,, 
\ee
where $\tilmu$ denotes the nonrelativistic chemical potential.
Equation~\eqref{eq:eom_phi2} multiplied by $1/\sqrt{2m}$ then becomes
\be 
  (\tillam_2 + \tilmu - e n_e A^0)\psi + \frac{\bD^2}{2m} \psi
  -\frac{\lambda_4}{4m^2} |\psi|^2 \psi = 0\,,
\label{eq:eom_psi}
\ee
where we have dropped a subleading term proportional to
$(\tilmu-e n_e A^0)^2/(2m)$.

We should solve Eq.~\eqref{eq:eom_psi} together with
Eqs.~\eqref{eq:eom_A} and \eqref{eq:eom_A0} for EM fields.  The Gauss law~\eqref{eq:eom_A0}
reads:
\be 
  \bnabla^2 A^0 + e n_e |\psi|^2 + q 
  = - \frac{e n_e}{m} (\tilmu - e n_e A^0) |\psi|^2 \simeq 0\,,
\ee
where we again drop the last term which is subleading according to the
approximation made in Eq.~\eqref{eq:eom_psi}, while we still keep the
kinetic term $\bD^2/(2m)$ in Eq.~\eqref{eq:eom_psi}.  For notational
brevity, let us rename our variables as follows:
\be
  \tillam_2 + \tilmu \;\to\; \mu\,,\qquad
  \frac{\lambda_4}{4m^2} \;\to\; g\,,\qquad
  A^0 = \frac{\mu}{e n_e} a\,,\qquad
  \psi \;\to\; \sqrt{\frac{\mu}{g}}\,\Psi\,.
\ee
Here, we note that this $\mu$ is different from the original one in
Eq.~\eqref{eq:eom_phi2}.  Together with the Maxwell equation for $\bA$
in Eq.~\eqref{eq:eom_A}, our equations finally become
\begin{align}
  & (1-a)\Psi + \frac{1}{m_H^2}(\bnabla - ie n_e \bA)^2 \Psi
    - |\Psi|^2 \Psi = 0\,,\\
  &\bnabla\times (\bnabla\times \bA)
    + m_V^2 \biggl[ \bA |\Psi|^2 -\frac{i}{2e n_e} \bigl( \Psi \bnabla\Psi^\dag
    - \Psi^\dag \bnabla\Psi \bigr) \biggr] = 0\,,\\
  &\bnabla^2 a + 2m^2 \frac{m_V^2}{m_H^2} (|\Psi|^2 + \tilq) = 0\,,
\end{align}
where $\tilq\equiv (g/e n_e \mu)q$, and we also introduce the two
typical mass scales as
\be
  m_H^2 \equiv 2 m \mu\,,\qquad
  m_V^2 \equiv \frac{(e n_e)^2 \mu}{m g}\,.
\ee
Physically, $1/m_H$ represents the coherent length of the field
$\Psi$, while $1/m_V$ represents the penetration length of the
magnetic field.  If the penetration length is smaller than the
coherent length, $m_V > m_H$, the Meissner screening effect is
dominant and the phase separation is more preferred than forming
magnetic vortices, which corresponds to Type-I superconductivity.  We
are interested in Type-II superconductivity in the opposite regime
with $m_H > m_V$.

The Ansatz for the vortex solution with the winding number $\nu$ is
\be
  \Psi = f(r)\, e^{i \nu \varphi}\,,\qquad
  a = a(r)\,,\qquad
  A^i = -\frac{\nu}{e n_e}\,\varepsilon^{ij}\,\frac{x^j}{r^2} \bigl[ 1-h(r) \bigr]\,,
  \label{eq:vortex}
\ee
where $r\equiv\sqrt{x^2+y^2}$ and $\tan\varphi\equiv y/x$.
Introducing a dimensionless radial coordinate,
$\rho\equiv m_V\, r$, we can rewrite the differential equations (with
$'\equiv {d\over d\rho}$) as
\begin{align}
  &-(\rho\, f')' + \frac{\nu^2 h^2}{\rho}\, f + \lambda\, \rho\, f (f^2 -1 + a) = 0\,,\\
  &\rho\biggl(\frac{h'}{\rho}\biggr)' - f^2 h = 0\,,\\
  &\frac{1}{\rho} (\rho\,a')' + \frac{2}{\lambda}\,\frac{m^2}{m_V^2} (f^2 + \tilq) = 0\,,
    \label{eq:aprime}
\end{align}
where $\lambda\equiv m_H^2/m_V^2>1$. 
For the total charge neutrality condition, we impose the condition,
\be
 \int_{\bx} \tilq = -\int_{\bx}\, f^2 \,.
\label{eq:tilq}
\ee
Here, $\int_{\bx}$ refers to the two-dimensional integration on the
plane perpendicular to the vortex axis.  This neutrality condition is
demanded by the fact that the static potential would behave as
$a(\rho\gg 1)={Q\over 2\pi}\log \rho$ if the total net charge $Q$ is
nonzero.  The combination of $(\mu-e n_e A^0)$ appears in the
equations of motion and it plays a role of an effective chemical
potential.  To have a well-defined effective chemical potential at
spatial infinity, we should impose $Q=0$.

\begin{figure}
  \includegraphics[width=0.47\textwidth]{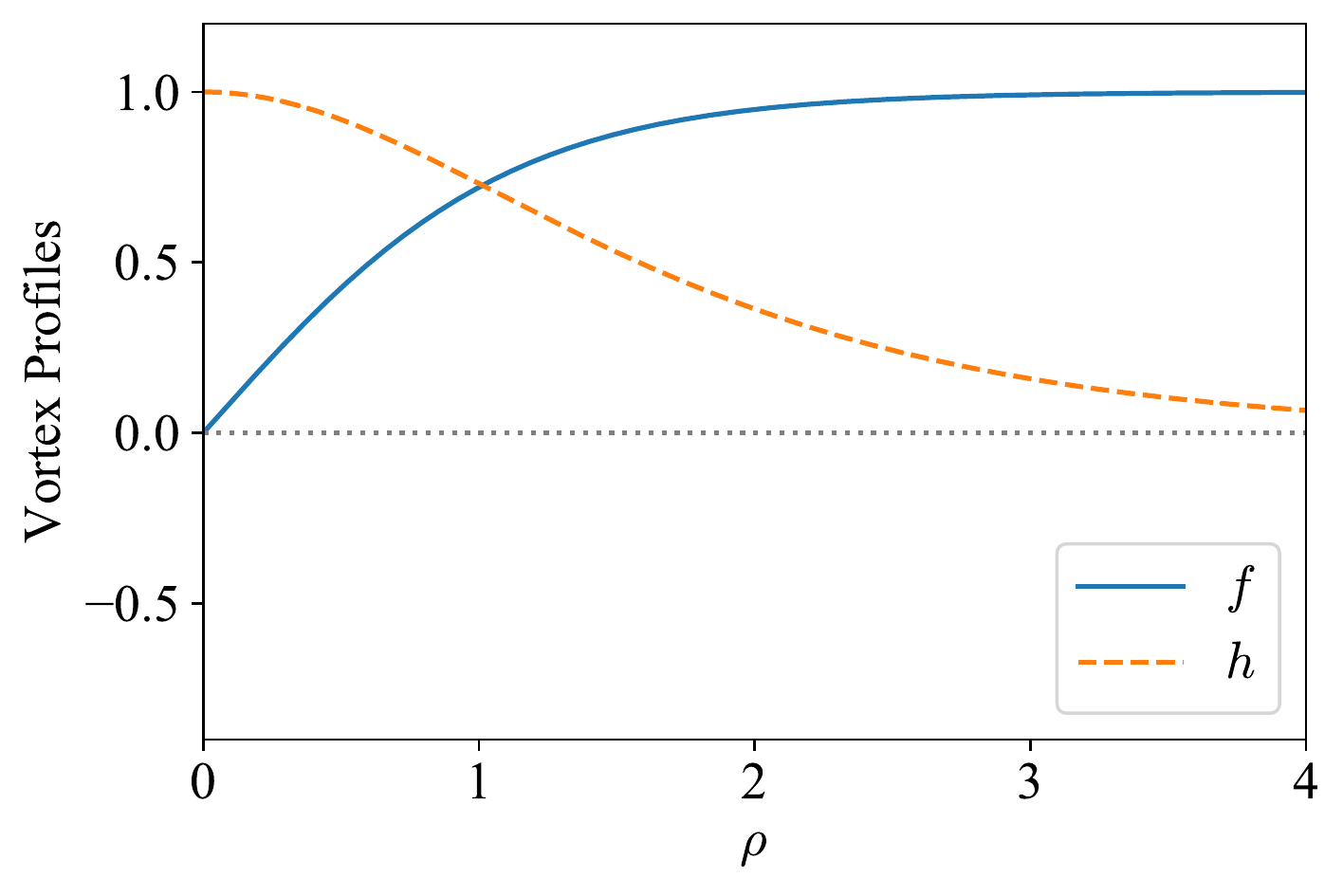} \hspace{1em}
  \includegraphics[width=0.47\textwidth]{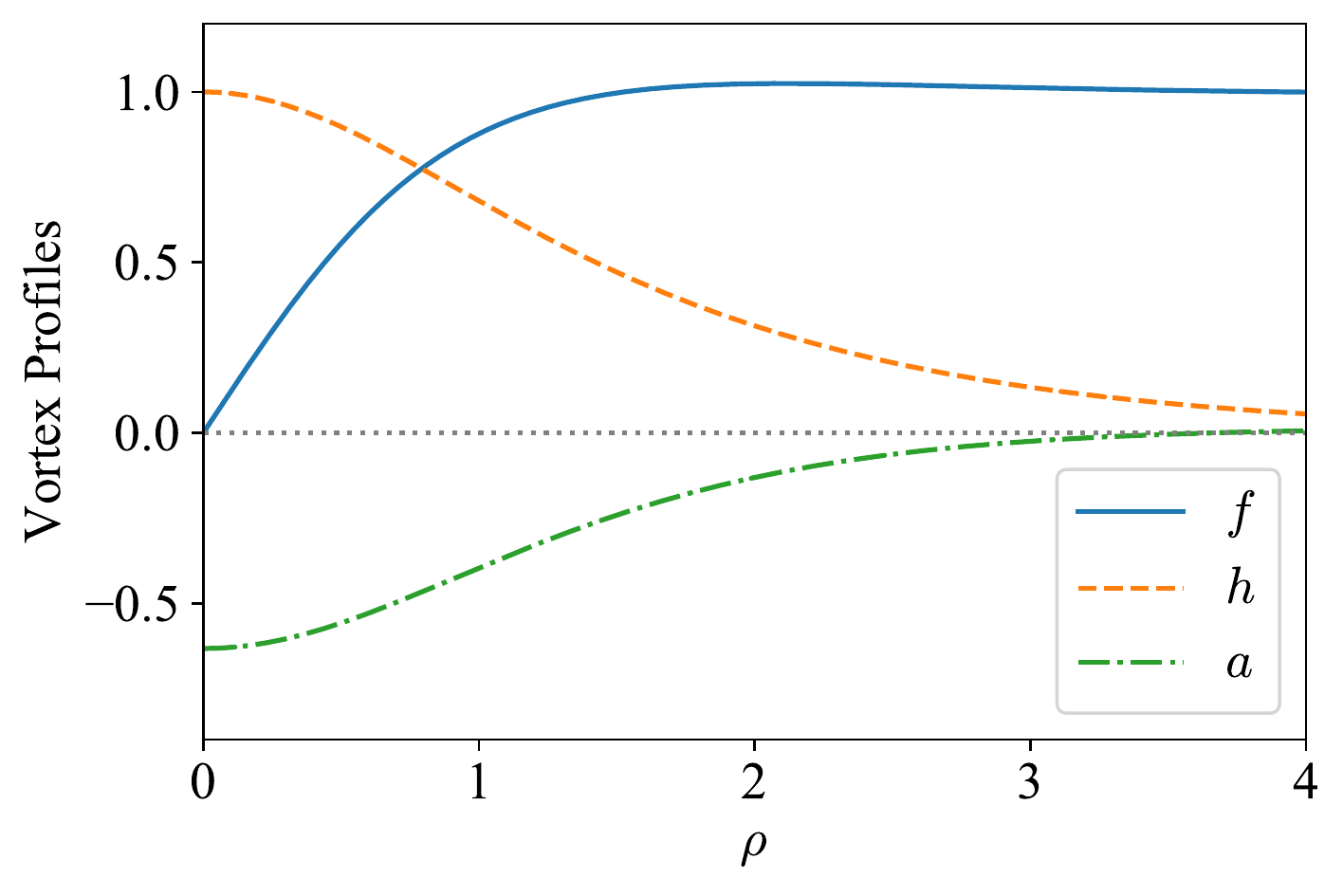}
  \caption{(Left panel) Profile of the conventional elementary ($\nu=1$) 
    magnetic vortex; $f$ and $h$ without coupling to $a$ for 
    $\lambda=1.5$.  (Right panel) Profile of the elementary vortex 
    with the electric field; $f$, $h$, and $a$ for $\lambda=1.5$ and 
    $m^2/m_V^2=1$.}
  \label{fig:vortex}
\end{figure}

We can numerically solve these differential equations with appropriate
boundary conditions.  Let us first consider the conventional ``locally
neutral'' vortex solution without coupling to electric field, so that
$a(r)= 0$ simply.  This can be achieved by choosing a space dependent
background charge density $\tilde q(x)$ that locally neutralizes the
net charge; that is, $f^2+\tilde q=0$, leading to $a(r)=0$ from
Eq.~\eqref{eq:aprime}.  Most Type-II vortices behave this way, but
there are examples where this does not happen in general; see
Refs.~\cite{PhysRevLett.75.1384,CLAYHOLD2003272}.
The regularity of $\Psi$ at $\rho=0$ requires $f(0)=0$, and at
infinity it should approach the vacuum value of $f(\infty)=1$.  In the
absence of $a$, then the boundary conditions should be
\be  
  f(0) = 0\,,\qquad  
  f(\infty) = 1\,,\qquad  
  h(0) = 1\,,\qquad  
  h(\infty) = 0\,.  
\ee
We can easily obtain the numerical solutions using the shooting method
to satisfy these boundary conditions.  The left panel of
Fig.~\ref{fig:vortex} shows an example of the profile of the magnetic
vortex for $\lambda=1.5$.  We see that $h(\rho)$ extends more widely
than $f(\rho)$, reflecting $m_H > m_V$.

As a nontrivial example where the local charge density and the
electric field are nonvanishing, let us consider a constant background
charge density $\tilde q$, that is determined by the total charge
neutrality condition~{\eqref{eq:tilq}} as
\be
  \tilq = -\frac{1}{S} \int_{\bx}\, f^2 \,,
  \label{eq:tilq2}
\ee
with $S\equiv \int_{\bx}$ is the transverse area.  In the limit of
infinitely large system $\tilde q$ would approach the negative unity.
In the present case we should revise the boundary conditions
accordingly.  That is, $f$ needs not be unity at large $\rho$, but
$f^2-1+a$ should be vanishing as $\rho$ gets large.  Also, we
physically require vanishing electric field at $\rho=0$ and
$\rho\to\infty$.  Therefore, we impose the following boundary
conditions:
\be 
f(0) = 0\,,\quad 
f(\infty) = \sqrt{1-a(\infty)}\,,\quad 
h(0) = 1\,,\quad 
h(\infty) = 0\,,\quad
a'(0)=0\,,\quad
a'(\infty)=0\,.
\label{eq:boundary}
\ee 
Actually, these boundary conditions are not sufficient to determine
the numerical solution uniquely, but a shift of
$a(\rho) \to a(\rho)+c$ with a constant $c$ still exists.  
This shift would change the value of $\mu$, and the magnitude of condensate would
also be modified, which would result in a different value of $\tilq$
in {Eq.~\eqref{eq:tilq2}.}  In other words, we can adjust $\tilq$ to make
a constant shift on $a(\rho)$.  To fix this freedom, a natural
condition to impose would be to set $a\to0$ at large $\rho$, so that
the effective chemical potential at infinity, by definition, remains to be $\mu$.
We choose $\lambda=1.5$ and $m^2/m_V^2=1$ to find that
$a(\infty)\to 0$ is realized with $\tilq\simeq -0.985$.  In the right
panel of Fig.~\ref{fig:vortex} we present the numerical solution with
these parameters.  This explicitly demonstrates that nontrivial
solutions with nonzero local charge density and electric field
certainly exist.  We see that the profile of condensate slightly
shrinks as compared to the locally neutral case shown in the left
panel.

Let us now compute the angular momenta carried by the matter and the
EM fields.  The matter part of the angular momentum per unit vortex
length is
\be 
  L^{\rm matter}_z = \int_{\bx} \psi^\dag
  \Bigl( \frac{\hbar}{i}\, D_\varphi \Bigr) \psi \,,
  \qquad D_\varphi\equiv \partial_\varphi - {ie n_e\over\hbar}\, A_\varphi\,,
  \qquad A_\varphi \equiv\epsilon^{ij}x^iA^j=\frac{\nu\hbar}{e n_e}[1-h(r)]\,,
\label{eq:Lzkin}
\ee
where we reinstate $\hbar$ as a common unit for the angular momentum
and also change the variables back to $r$ and $\varphi$.  We note that
the boundary condition~\eqref{eq:boundary} guarantees
$D_\varphi [f(r) e^{i\nu\varphi}] \to 0$ as $r\to \infty$, and the
above integral is convergent.  It should be mentioned that the above
form of the angular momentum using $D_\varphi$ corresponds to the
\textit{kinetic angular momentum}, that is the angular momentum
carried by matter alone.  We could have defined the
\textit{canonical angular momentum} using $\partial_\varphi$.  It is
straightforward to find:
\be
  L^{\rm can, matter}_z = \int_{\bx} \psi^\dag
  \Bigl( \frac{\hbar}{i}\, \partial_\varphi \Bigr) \psi =
  \nu (2\pi\hbar) \frac{\mu}{g}\int_0^R dr\, r\, f^2(r)
  = \nu\hbar\,N\,,
\label{eq:Lcan}
\ee
where $N\equiv\frac{\mu}{g}\int_{\bx} f^2$ is the total number of
particles per unit vortex length, and $R$ is the size of the system in
radial direction.  This expression is identical to the well-known one
for the quantized angular momentum of a superfluid vortex.  
In a gauge theory, there is a contribution from the gauge field:
\begin{equation}
L^{\rm can, gauge}_z=\int_{\bx}\bigl[\bE\cdot \partial_\varphi\bA+(\bE\times\bA)_z\bigr]\,,
\end{equation}
which vanishes in the vortex configuration~\eqref{eq:vortex}.
The sum of $L^{\rm can, matter}_z$ and $L^{\rm can, gauge}_z$ gives
the total angular momentum $L^{\rm can}_z$ which is conserved.
Alternatively we can consider the conserved
angular momentum as the sum of $L^{\rm matter}_z$ and the EM
contribution, $L^{\rm gauge}_z$, i.e.,
$L^{\rm total}_z = L^{\rm matter}_z + L^{\rm gauge}_z$.  
Which of
$L^{\rm can}_z$ or $L^{\rm total}_z$ is the relevant angular
momentum depends on the physical setup.  In our present setup we can
gradually turn on the magnetic field, so that the magnetic vortex
emerges.  In this case, it makes sense to consider
$L^{\rm total}_z$, not $L^{\rm can}_z$.  
Although the difference between $L^{\rm can}_z$ and $L^{\rm total}_z$ is the only boundary term,
it plays an essential role in the conservation of angular momentum as
discussed in Sec.~\ref{sec:example 1}.
For more discussions on the canonical angular momentum, see a concrete
analysis in Ref.~\cite{Greenshields:2014caa} and also a general
consideration in Ref.~\cite{Leader:2015vwa}.

\begin{figure}
  \includegraphics[width=0.47\textwidth]{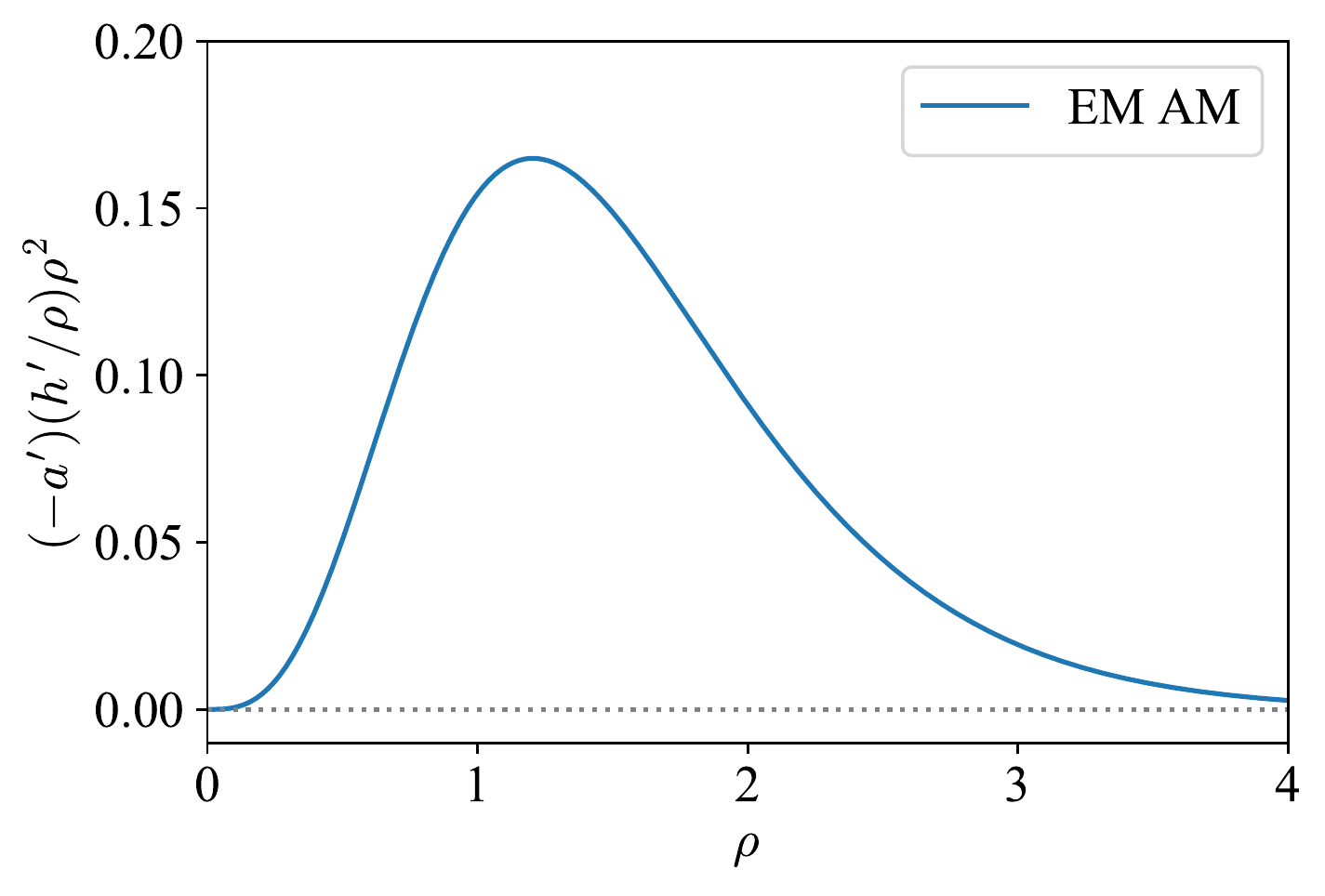}
  \caption{The integrand of Eq.~\eqref{eq:LEM2} in terms of  
    dimensionless variables for $\lambda=1.5$ and $m^2/m_V^2=1$, which
    represents the local distribution of the EM angular momentum.}
  \label{fig:EM_AM}
\end{figure}

With the explicit forms of the vortex profile and the associated
vector potential, the matter part of the angular momentum per unit
vortex length becomes
\be 
L^{\rm matter}_z = \nu (2\pi\hbar) \frac{\mu}{g}
\int_0^R dr\, r\, h(r)\, f^2(r)\,.\label{eq:Lmatter1}
\ee 
The difference from the canonical expression~\eqref{eq:Lcan} is the
presence of $h(r)$ in the integrand.  Because $h(r)$ decays when $f(r)$
increases as in Fig.~\ref{fig:vortex}, we see that $L^{\rm matter}_z$
is smaller than $L^{\rm can}_z$.  Let us next consider the EM
contribution, i.e.,
\be 
  L^{\rm gauge}_z = \int_{\bx}  \bigl[ \bx\times(\bE \times \bB) \bigr]_z\,.
\label{eq:LEM}
\ee 
Plugging the explicit forms of $\bE$ and $\bB$ into the above, we find
$L^{\rm gauge}_z$ as
\be 
  L^{\rm gauge}_z = -(2\pi\hbar)\frac{\nu\mu}{(e n_e)^2} \int_0^R dr\,
\biggl[ r\frac{da(r)}{dr}\biggr]\, \frac{dh(r)}{dr}\,. 
\label{eq:LEM2}
\ee
In Fig.~\ref{fig:EM_AM} the integrand corresponding to the local
angular momentum density is plotted, where the variables are made
dimensionless again.

The angular momentum distribution is peaked around $\rho\sim 1$ and
decays at large $\rho$.  We can perform an integration by part and use
the equation of motion to transform the above expression into
\be
  L^{\rm gauge}_z = -(2\pi\hbar)\frac{\nu\mu}{(e n_e)^2} \biggl\{
  r \frac{da(r)}{dr} h(r) \biggl|_0^R + 2\, \frac{m^2}{\lambda}
  \int_0^R dr\,r\,h(r) \bigl[ f^2(r) + \tilq(r) \bigr] \biggr\}\,.
\ee
Because of the boundary conditions~\eqref{eq:boundary}, the surface
contribution vanishes.  Using $\lambda=m_H^2/m_V^2$ we can simplify
the above expression into
\be
  L^{\rm gauge}_z = -\nu(2\pi\hbar)\frac{\mu}{g} \int_0^R dr\, r\,
  h(r)\,\bigl[ f^2(r) + \tilq(r) \bigr]\,.
\ee
Comparing with the matter contribution $L_z^{\rm matter}$ in
Eq.~\eqref{eq:Lmatter1}, the first term is remarkably equal to
$-L_z^{\rm matter}$, and the total kinetic angular momentum is thus,
\be
  L^{\rm total}_z = \, -\nu (2\pi\hbar) \frac{\mu}{g}
\int_0^R dr\,  r\, h(r)\tilq(r)\,.
\ee
We see that $L^{\rm  total}_z$ is proportional to $\tilq$ and this
nonzero value of the total angular momentum is attributed to the
presence of the background.  
If we had no background, $\tilq=0$, then $L^{\rm matter}_z$ and
$L^{\rm gauge}_z$ would have perfect cancellation, but we then allow for
a ``charged'' magnetic vortex.  This might be possible due to
finiteness of the system bounded by $R$.  A natural realization of
this possibility will be discussed as Class~II in the next section.

\subsection{Class Ib -- Incomplete cancellation due to topological boundary}

Our next example for incomplete cancellation has been motivated by the
physical setup discussed in Ref.~\cite{PhysRevLett.121.227001} where a
fractional angular momentum in the units of $\hbar$ is found to be
carried by the magnetic vortex at the interface between a
superconductor and a $T$-invariant strong topological insulator
(TI)\footnote{We note that our result derived in the following is
  different by a factor $1/2$ from Ref.~\cite{PhysRevLett.121.227001}.
  We have identified where this difference stems from, but it is not
  essential for our present argument, so we will not go into that detail.}.
We will consider a simplified situation that still demonstrates the
essential physics involved;  we will show that a nonzero and fractional
angular momentum is localized around a magnetic flux on the boundary
surface between a bulk TI and the vacuum outside.  Let us think of
this situation from a different perspective.  In the same way we
discussed Class~Ia in the previous section, we can imagine a procedure
to turn on the magnetic field gradually from zero, and yet the angular
momentum conservation guarantees zero total angular momentum of the
whole system.  The only way our result of fractional angular momentum
can be consistent with the angular momentum conservation is that the
other compensating angular momentum should be located in the other
part of the TI-vacuum boundary where the magnetic flux leaves out from
the bulk TI.  If this boundary region is far separated from the
place where the original incoming flux enters the TI, we can
reasonably neglect this far-away region, and focus only on the angular
momentum localized on the incoming flux alone.  This angular momentum
indeed takes a fractional value, as we confirm in the following
discussions.  We can say that the fractional angular momentum is
transported from the boundary at infinity to the incoming magnetic
flux; this characterizes the magnetic vortices of Class~Ib in our
classification.  Such magnetic vortices with fractional angular
momenta are not peculiar, but rather ubiquitous in topological phases
of matter;  see, for example,
Refs.~\cite{Seiberg:2016rsg,Seiberg:2016gmd,Senthil:2018cru}.  A
deeper insight to the angular momentum conservation from our
discussion should be useful for better understanding of these
systems.

\begin{figure}
  \includegraphics[width=0.35\textwidth]{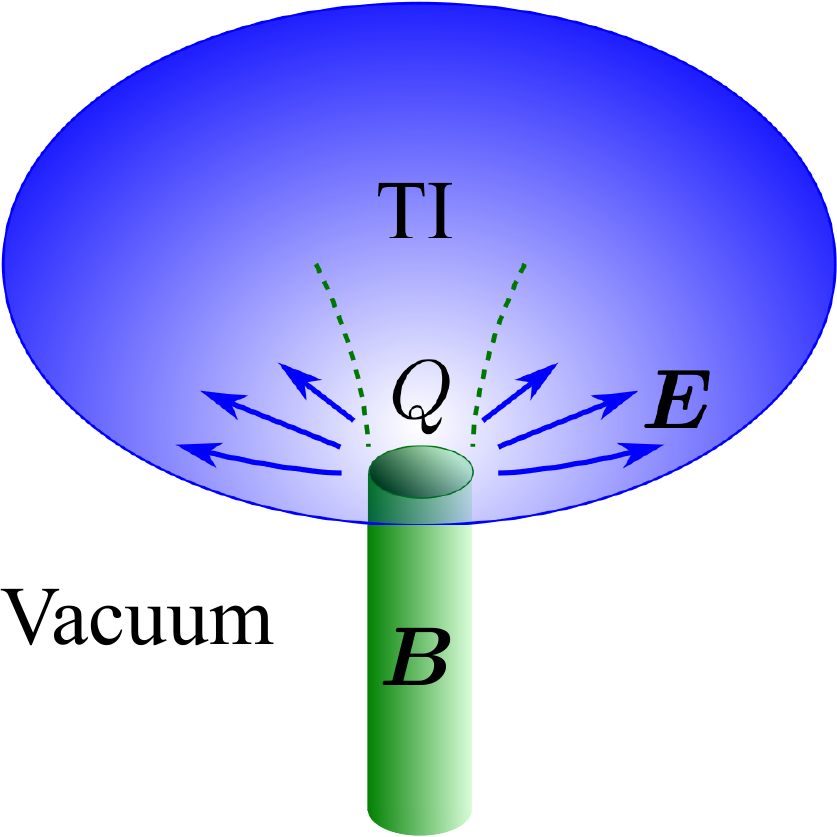}
  \caption{Interface between the TI and the vacuum with a localized 
    flux of magnetic field $\bB$.  The electric charge $Q$ is stored at 
    the interface which produces the electric field $\bE$.}
  \label{fig:TI}
\end{figure}

Let us consider a situation where we have a TI bulk in the $z>0$
region and the vacuum in $z<0$, with an interface at the $z=0$
surface, as illustrated in Fig.~\ref{fig:TI}.  It is well known that
the boundary of TI supports massless surface states that can be
described by a single Dirac fermion field.  For our purpose, let us
assume that there are $T$-violating magnetic impurities on the
surface, that opens a mass gap for the surface fermions.  Integrating
out the massive surface fermion gives rise to a new term in the
effective action in the low energy limit for the EM fields, which is
the Chern-Simons action with a half integer level,
$\nu={1\over 2}$~\cite{Qi:2008ew}  (which should not be confused with
the winding number in the previous subsection).  To capture the
essential physics of our discussion, we will consider an idealized
situation that this is the only response of the TI surface (with
$T$-violating impurities) to an externally applied EM field.  At least
in long wavelength and time limit, the Chern-Simons term becomes
dominant over other higher derivative terms in the action.

From the Chern-Simons action, the charge current in response to an
applied EM field is obtained as
\be 
  j^\mu = -{\nu\over 2}{e^2\over h }\,\epsilon^{\mu\nu\alpha}
  F_{\nu\alpha}
  = -{e^2\over 8\pi\hbar }\,\epsilon^{\mu\nu\alpha}F_{\nu\alpha}\,,
\ee
which in components reads as 
\be 
  Q = {e^2\over 4\pi\hbar} B_z\,,\qquad
  j_x = {e^2\over 4\pi\hbar} E_y\,,\qquad
  j_y = -{e^2\over 4\pi\hbar} E_x\,.
  \label{eq:QJ}
\ee
where $Q$ and $j_{x,y}$ are the charge density and the quantum Hall
effect (QHE) current, respectively.  Here, $j_{x,y}$, $B_z$, and
$E_{x,y}$ in Eq.~\eqref{eq:QJ} represent 3D vector components without
distinction between upper and lower indices.

We consider a magnetic flux that is vertically piercing the interface
and is cylindrically symmetric: $\bB = B_z(r)\hat{z}$, where $r$ is the
radius in the $x$-$y$ plane.  We further assume that $B_z(r)$ is
localized for $r\le R$, so we can regard it as a flux tube like a
magnetic vortex.  In fact, we may realize such a magnetic profile by
an external superconducting vortex as postulated in
Ref.~\cite{PhysRevLett.121.227001}.  As seen from Eq.~\eqref{eq:QJ}
the magnetic flux induces a surface charge density $Q$ and this charge
gives rise to a nonzero electric field according to the Gauss law.  It
is easy to understand that a nonzero angular momentum emerges from
the resulting EM fields which are indicated by arrows in
Fig.~\ref{fig:TI}.

Before going into the computation of the angular momentum carried by
the EM field, we first show that the angular momentum contribution
from the TI matter part at $z>0$ is generally vanishing.  The easiest
way to confirm this is to compute the angular momentum that may be
transferred to the TI surface states as we increase the magnetic flux
from zero.  This is because the TI bulk is gapped, and only the
surface states may carry angular momentum in response to the applied
EM fields in the system.  During the process of turning on the
magnetic flux, we have a tangential electric field $E_\varphi=(xE_y-yE_x)/r$ from
Faraday's law,
\be
  2\pi r E_\varphi(r)
  = -2\pi \int_0^r dr' \,r'\, {\partial B_z(r',t)\over \partial t}\,.
\ee 
Then, according to Eq.~\eqref{eq:QJ} in the cylindrical coordinates,
we have the QHE current as $j_r = (xj_x+yj_y)/r ={e^2\over 4\pi\hbar} E_\varphi$.
From this, we can compute the torque from the EM force acting on the
surface states along the $\varphi$ direction.  The EM force reads,
\be 
  F_\varphi = Q E_\varphi - j_r\,  B_z\,,
\ee 
where the second term represents the Lorentz force of
$\bj \times \bB$.  Using $Q={e^2\over 4\pi\hbar}B_z$ and
$j_r={e^2\over 4\pi\hbar} E_\varphi$, we see that the force vanishes
identically, that is, the surface states do not experience any
tangential force, or torque, by the Chern-Simons term.  In fact, it is
easy to verify that this result generally holds for any geometry.  We
conclude that no angular momentum is carried by the TI matter and its
surface states.  The angular momentum of the whole system resides
solely in the EM sector.

Now let us return back to the angular momentum in the EM sector.  For
static fields satisfying $\bnabla\times\bE=\boldsymbol{0}$ and the
vector potential $\bA=A_\varphi \hat{\varphi}/r$ satisfying
$\bnabla\cdot\bA=0$, we can rewrite the angular momentum of EM fields
as
\be 
  L_z = \int_{\bx} \bigl[ \bx \times (\bE \times\bB) \bigr]_z
  = \int_{\bx} (\bnabla\cdot\bE)\,  A_\varphi
  - \int   A_\varphi (\bE \cdot d\boldsymbol{S})\,, 
\label{eq:Lzseparate}
\ee 
where the last term is the surface integral on the exterior boundary.
Using the Stokes theorem and the cylindrical symmetry, we find the
vector potential with the boundary condition, $A_\varphi(0)=0$, to be
\be 
  2\pi  A_\varphi(r) = 2\pi\int_0^r dr'\, r' B_z(r')\,.
\ee 
Then, with the Gauss law, $\bnabla\cdot\bE=Q\delta(z)$, the first term
in the above expression of $L_z$ becomes
\be 
  \int_{\bx} (\bnabla\cdot\bE)\,  A_\varphi
  = {e^2\over 2\hbar} \int_0^\infty dr\, r B_z(r)\int_0^r dr'\, r' B_z(r')
  = {e^2\over 4\hbar}\biggl[ \int_0^\infty dr\, r B_z(r) \biggr]^2
  = {e^2\over 16\pi^2\hbar} \Phi_0^2\,,
\ee 
where $\Phi_0$ is the total magnetic flux.  For the second term, we
consider a cylindrical boundary at $r=R$, and the Stokes theorem leads
to
\be 
  2\pi  A_\varphi(R) = \Phi_0\,,
\ee 
which takes a constant value along the boundary.  Then, the vector
potential can be taken out from the integrand, which gives
\be 
  -\int\, A_\varphi (\bE\cdot d\boldsymbol{S})
  = -{\Phi_0\over 2\pi}\int \bE\cdot d\boldsymbol{S}
  = -{\Phi_0\over 2\pi}Q_{\rm tot}
  = -{e^2\over 8\pi^2\hbar}\Phi_0^2\,,
  \label{eq:2ndL}
\ee 
where $Q_{\rm tot}={e^2\over 4\pi\hbar}\Phi_0$ from Eq.~\eqref{eq:QJ}
is used.  Summing the above two terms, we get the total angular
momentum as
\be 
  L_z = -{e^2\over 16\pi^2\hbar} \Phi_0^2 
\ee 
with the right sign that can easily be confirmed.  We note that the
original integral of the angular momentum is convergent by itself as
$B_z(r)$ is of finite range, and the above way to split it into two
terms is just a mathematical manipulation for convenience.

We shall suppose that the magnetic flux is quantized as if it were
provided by an adjacent superconducting vortex of the winding number
$\nu$ considered in Ref.~\cite{PhysRevLett.121.227001}.  We note that
the magnetic vortex in superconductivity does not carry a finite net
angular momentum except for the background contribution, so that the
total angular momentum of our interest is still given by the above
formula.   The flux quantization gives
${2e\over\hbar }\Phi_0=2\pi \nu$, where the factor 2 of
$\frac{2e}{\hbar}$ originates from the Cooper pair.  This finally
leads to
\be 
  L_z = -{\nu^2\over 16}\hbar\,.
  \label{eq:TILz}
\ee 
Therefore, the EM field surrounding the magnetic vortex between a TI
and a superconductor carries a nonzero angular momentum given in
Eq.~\eqref{eq:TILz}.

The conservation of the total angular momentum during the process of
turning on the magnetic flux requires the existence of an opposite and
compensating angular momentum somewhere else.  To identify where this
compensating component is, let us consider a global geometry of the
bulk TI and its closed boundary.  For simplicity we assume that the
bulk TI (which is a blue shaded region in Fig.~\ref{fig:TI}) is a
large ball of radius $R$ and the boundary surface is a sphere of
radius $R$.  A localized magnetic tube with a flux $\Phi_0$ enters the
TI at $\theta=\pi$, where $\theta$ is the polar angle in 3D
spherical coordinates.  The same flux leaves out from the TI at other
places of the surface in cylindrically symmetric (i.e., $\varphi$
independent) way.  Let the radial component of the magnetic field at
$r_{\rm 3D}=R$ be $B_{r_{\rm 3D}}(\theta)$ as a function of $\theta$,
where $r_{\rm 3D}^2=r^2+z^2$ is the 3D radius. 
 The flux
conservation results in
\be 
  \int_0^\pi d\theta\, \sin\theta\, B_{r_{\rm 3D}}(\theta) = 0\,.
  \label{eq:zeroB}
\ee
We consider turning on the magnetic field adiabatically from zero, and
the time-dependent magnetic field gives $E_\varphi$ as well as the QHE current
$j_\theta$, but the net force on the surface states is vanishing as we
saw before.  Therefore, the total angular momentum resides in the EM
fields only.

To compute the EM part of the angular momentum, we follow the same
steps as before.  Previously we considered only the contribution from
the incoming magnetic flux, but if we perform the same computation
including the whole TI boundary surface, the total $L_z$ turns out to
be zero as we show in the following, that is consistent with our
angular momentum conservation argument.  From the spherical symmetry
of the TI geometry and the cylindrical symmetry of the vector
potential, we have
\be 
  2\pi  A_\varphi(\theta) = 2\pi R^2\int_0^\theta d\theta'\,\sin\theta' \,B_{r_{\rm 3D}}(\theta')\,.
\ee
The Gauss law gives
\be 
  \bnabla\cdot\bE = {e^2\over 4\pi\hbar}B_{r_{\rm 3D}}(\theta)\delta(r_{\rm 3D}-R)\,.
\ee 
Then, we find the
first term in Eq.~\eqref{eq:Lzseparate} to be
\begin{align}
  \int_{\bx}(\bnabla\cdot\bE)\, A_\varphi
  &= 2\pi R^4 {e^2\over 4\pi\hbar} \int_0^\pi\,d\theta\, \sin\theta\,
  B_{r_{\rm 3D}}(\theta) \int_0^\theta\,d\theta' \,\sin\theta'\, B_{r_{\rm 3D}}(\theta') \notag\\
  &= 2\pi R^4 {e^2\over 4\pi\hbar} \,{1\over 2}
  \biggl[ \int_0^\pi\,d\theta\, \sin\theta\, B_{r_{\rm 3D}}(\theta)\biggr]^2=0\,,
\end{align}
using Eq.~\eqref{eq:zeroB}.  For the second term in
Eq.~\eqref{eq:Lzseparate}, we can still employ Eq.~\eqref{eq:2ndL}
with different $Q_{\rm tot}$.  Previously we took account of
$Q_{\rm tot}$ around the incoming magnetic flux only, but if we sum up
all the contributions from the whole TI surface, it should amount to
$Q_{\rm tot}=0$ due to Eq.~\eqref{eq:zeroB}.  In this way we see that
the second term is zero as well.  We emphasize that the original
expression of the angular momentum is localized in the region where
$\bB\neq \boldsymbol{0}$ and $\bE\neq \boldsymbol{0}$, that is, it is
localized around the TI boundary where the magnetic flux either enters
or leaves the TI.  Therefore, the fractional angular momentum
localized around the magnetic tube at $\theta=\pi$ is compensated by
the angular momentum carried by the outgoing flux in other places of
the TI boundary which can be taken infinitely away.

\section{Class II: Case study of magnetic vortices with zero angular
  momentum}

In this section we consider magnetic vortices in relativistic field
theory as examples of self-consistent systems without any background
matter or boundary that could absorb angular momentum.  Such vortices
could appear in the Standard Model and extensions of the Standard
Model.  They have been considered in the context of high energy
physics and cosmology.  A faithful application of our angular momentum
conservation argument to these vortices then dictates that they should
be spinless.  We will confirm this claim also in a nontrivial example
where the angular momentum carried by surrounding localized gauge
fields is essential for the cancellation of the total angular
momentum.  We emphasize that these localized gauge fields around the
vortex core should be considered as a part of the magnetic vortex
configuration under consideration.

\subsection{Example 1: Relativistic Nielsen-Olesen vortices}
\label{sec:example 1}

Let us illustrate our main points in the simplest example of
Nielsen-Olesen vortices in relativistic scalar theory that we already
treated in the previous section.  The formulation of the theory presented
below is a standard one, but we would like to pay a special attention
to the cases with nonvanishing charge density.  Consequently, nonzero
electric fields accompany the vortices; we then have a precise
description of the charged vortices, taking proper account of the
Gauss law constraint and the Coulomb energy contribution to the
Ginzburg-Landau free energy to be minimized.  This endeavor, that we
did not find in the literature in full generality as we present here,
turns out to be crucial to show the exact cancellation of total
angular momenta carried by the matter and the gauge field parts.

This section has some redundancy with our discussions in the previous
section, but to make our analysis as self-contained as possible,
let us retain some calculational details.  Showing explicit terms for our
later convenience, we write down the Lagrangian as
\be 
  \calL = (D_0\Phi)^\dagger (D_0\Phi)-(\bD\Phi)^\dagger (\bD\Phi)
  -U(\Phi^\dagger\Phi) + {1\over 2}\bE^2-{1\over 2}\bB^2\,,
\ee
which is Eq.~\eqref{eq:Lscalar} without background, i.e., $q=0$.
Here, we take $n_e=1$ and
$D_\mu \Phi=(\partial_\mu + ieA_\mu) \Phi$ and, as defined in
Sec.~\ref{sec:ClassIa}, we adopt a convention of $(\bD)^i=D^i$.
The EM fields are $\bE=-\bnabla A_0-\partial_0 \bA$ and
$\bB=\bnabla\times\bA$.  We use the unit system with $c=\hbar=1$ in
this section.  We also take a conventional form of the potential same
as in the previous section;
$U(\Phi^\dagger\Phi)=-\lambda_2 \Phi^\dagger\Phi+
{\lambda_4\over 2}(\Phi^\dagger\Phi)^2$.
We reproduce the equations of motion and the Gauss law from this
action as
\begin{align}
  & -D_0^2\Phi+\bD^2\Phi- U'(\Phi^\dagger\Phi)\Phi=0\,,
    \label{eom}\\
  & -{\partial_0\bE}+\bnabla\times\bB
    +ie \bigl[ (\bD\Phi)^\dagger \Phi-\Phi^\dagger (\bD\Phi)\bigr]=0\,, \\
  & \bnabla\cdot\bE + ie\bigl[ (D_0\Phi)^\dagger \Phi
    -\Phi^\dagger(D_0\Phi)\bigr]=0\,,
\end{align}
which are equivalent to Eqs.~\eqref{eq:eom_phi}-\eqref{eq:eom_A0}
with $q=0$ and $n_e=1$.  We will work in the Hamiltonian formulation of the theory to look into the
dynamics further.

The canonical conjugate field is given by definition as
\be 
\Pi^\dagger ={\delta {\cal L}\over\delta \partial_0\Phi}=(D_0\Phi)^\dagger\,.
\ee 
It should be noted that in our convention the above expression defines
$\Pi^\dag$, not $\Pi$.  The charge density from the N\"{o}ther method
is
\be 
  Q = -i\bigl[ (D_0\Phi)^\dagger \Phi-\Phi^\dagger
  (D_0\Phi)\bigr] = -i(\Pi^\dagger\Phi-\Phi^\dagger\Pi)\,,
\ee 
and the Gauss law takes the form of
\be 
  \bnabla\cdot\bE = eQ = -i e(\Pi^\dagger\Phi-\Phi^\dagger\Pi)\,. 
\ee 
The Hamiltonian density from the Legendre transformation (including
the EM sector) is obtained as
\begin{align}
  H &= \Pi^\dagger (\partial_0\Phi)+\Pi(\partial_0\Phi)^\dagger
      -\bE (\partial_0 \bA) - \calL \notag\\
    &= \Pi^\dagger\Pi + (\bD\Phi)^\dagger (\bD\Phi)
      +U(\Phi^\dagger \Phi)+{1\over 2}(\bE^2+\bB^2)-ie A_0(\Pi^\dagger
      \Phi-\Phi^\dagger \Pi) -A_0 \bnabla\cdot\bE \notag\\
    &= \Pi^\dagger\Pi + (\bD\Phi)^\dagger (\bD\Phi)
      +U(\Phi^\dagger \Phi)+{1\over 2}(\bE^2+\bB^2)\,,
\end{align} 
where we dropped the total derivative term $\bnabla\cdot(A_0\bE)$ in
the second line, and from the second to the last line, we used the
Gauss law to have cancellation between the last two terms.  This
should be the case since $A_0$ is not a dynamical degrees of freedom
in the Hamiltonian formulation of gauge theory.

For our convenience we introduce a chemical potential $\mu$ via the
free energy to be minimized; $F=H-\mu Q$.  This is equivalent to
introducing $\mu$ in the covariant derivative, once $F$ in this
section is identified as the Hamiltonian density in the
previous section.
The free energy is explicitly given by
\be
  F = H-\mu Q = \Pi^\dagger\Pi+(\bD\Phi)^\dagger (\bD\Phi)
      +U(\Phi^\dagger \Phi)+{1\over 2}(\bE^2+\bB^2)
      +i\mu(\Pi^\dagger \Phi-\Phi^\dagger \Pi)\,.
\ee 
This, together with the Gauss law, constitutes a precise formulation
of gauged Ginzburg-Landau description for the cases with nonzero
charge distributions.  From the Gauss law, we see that $\bm E$ is not
independent but generated through $\Pi$ and $\Phi$, albeit in a
nonlocal way.  The variables, $\Pi$, $\Phi$, and $\bA$, are considered
as independent degrees of freedom, with respect to which the free
energy $F$ should be extremized to obtain the equations of motion.

We are interested in the stationary configurations where magnetic
field is static; $\partial_0\bm B=0$.
In this case, as is familiar in classical electromagnetism, we can
introduce an auxiliary function or static potential $A_0$ such that
$\bE=-\bnabla A_0$ and, with a proper boundary condition at spatial
infinity, the Gauss law can be solved nonlocally as
\be 
A_0=ie {1\over\bm\nabla^2}(\Pi^\dagger\Phi-\Phi^\dagger \Pi)\,,
\qquad A_0(\infty)=0 \,. \label{A0}
\ee 
This boundary condition is necessary, since a nonzero $A_0(\infty)$
would shift our definition of chemical potential $\mu$, that is, the
true chemical potential is $\mu-eA_0(\infty)$, as we have already seen
in the previous section.  Using this, one of the terms in $F$, that is, the electric field energy is expressed as 
\be 
{1\over 2}\bm E^2={e^2\over 2}(\Pi^\dagger\Phi-\Phi^\dagger \Pi) {1\over\bm\nabla^2}(\Pi^\dagger\Phi-\Phi^\dagger \Pi)\,,\label{gauss}
\ee 
which is nothing but the Coulomb energy induced by the charge
distributions.  The resulting expression for the free energy $F$
involves only the independent variables, $\Pi$, $\Phi$, and $\bm A$,
from which we can proceed to obtain the equations of motion.

From the variation with respect to $\Pi^\dagger$, we get
\be 
\Pi+i\mu\Phi+e^2\Phi  {1\over\bm\nabla^2}(\Pi^\dagger\Phi-\Phi^\dagger \Pi)=0\,. \label{Pi}
\ee 
Using the expression for $A_0$, this can be written as 
\be 
\Pi+i\mu\Phi-ieA_0\Phi=0
\qquad\text{or}\qquad \Pi=-i(\mu - eA_0)\Phi\,.
\ee 
Recalling the relation, $\Pi=D_0 \Phi=(\partial_0+ieA_0)\Phi$, this
gives the well-known Josephson relation;
\be 
\partial_0 \Phi=-i\mu\Phi\,.
\ee

Since $F$ is quadratic in $\Pi$, one may choose to insert back the
solution for $\Pi$ from Eq.~\eqref{Pi} into $F$ to get a more
conventional form of the free energy in terms of $\Phi$ and $\bA$
only.  It is explicitly given by
\be 
F=(\bm D\Phi)^\dagger (\bm D\Phi)+U(\Phi^\dagger \Phi)-\mu(\mu-eA_0)\Phi^\dagger\Phi +{1\over 2}\bm B^2 \,.
\ee 
One should keep in mind that $A_0$ in the above expression is a
functional of $\Phi$ that should be obtained by solving the Gauss
law~\eqref{A0} together with Eq.~\eqref{Pi}, which is in general
nonvanishing for $\mu\neq 0$ corresponding to nonzero charge
distributions in a solution.  A more practical way to approach this
problem is indeed what we have described in the preceding paragraphs,
i.e., keeping $\Pi$ as an independent degree of freedom. 

The variation of the free energy with respect to $\Phi^\dagger$ gives
\be 
-\bm D^2\Phi+\Phi U'(\Phi^\dagger \Phi)-i\mu\Pi-e^2\Pi{1\over\bm\nabla^2}(\Pi^\dagger\Phi-\Phi^\dagger\Pi)=0\,,
\ee 
which, upon using the expression for $A_0$, is equivalent to 
\be
-\bm D^2\Phi+ U'(\Phi^\dagger \Phi)\Phi-i(\mu - eA_0)\Pi=0 \,,
\ee 
and using the solution for $\Pi$, we finally get to 
\be 
-\bm D^2\Phi+ U'(\Phi^\dagger \Phi)\Phi-(\mu - eA_0)^2\Phi=0\,.\label{eq1}
\ee 
This, as it should, agrees with the equation of motion for $\Phi$
obtained from the Lagrangian, as written in Eq.~\eqref{eom}, after
using the Josephson relation, $\partial_0\Phi=-i\mu\Phi$.  However, we
again need to recall that $A_0$ is a solution of the Gauss law, which
itself involves $\Pi$ and $\Phi$ as
\be 
\bnabla^2 A_0=ie(\Pi^\dagger\Phi-\Phi^\dagger\Pi)
=-2e(\mu - eA_0)\Phi^\dagger\Phi\,.\label{eq2}
\ee 
Equations~\eqref{eq1} and \eqref{eq2} together with the Maxwell
equation for $\bm A$, i.e.,
\be 
\bnabla\times(\bnabla\times\bA) +ie\bigl[(\bD\Phi)^\dagger \Phi-\Phi^\dagger (\bD\Phi)\bigr]=0 \,,
\ee 
constitute our final set of closed equations for $\Phi$, $\bA$ and,
$A_0$ to be solved for classical configurations in relativistic
theory.  If $\mu=0$, then it is consistent with $A_0=0$, and there is
no electric field.  This situation at $\mu=0$ corresponds to charge
neutral vortices in our problem.  For $\mu\neq 0$ there exists
nonvanishing charge and the electric field in the solution, which we
will refer to as ``charged Nielsen-Olesen vortices.''

We could estimate the angular momentum in the matter part using an 
expression like Eq.~\eqref{eq:Lzkin}, but here, we shall show an 
alternative physical approach.  To compute the matter contribution to 
the angular momentum, we need the linear momentum density, i.e., 
$\bP$, obtained from $T^{0i}$ component of the energy-momentum 
tensor.  From the N\"{o}ther method, we see that $T^{0i}$ is given by 
\be 
\bP^i=T^{0i}=\bigl[(D^0 \Phi)^\dagger (\bD^i \Phi)+(\bD^i \Phi)^\dagger (D^0\Phi)\bigr]
=\bigl[\Pi^\dagger (\bD^i \Phi)+(\bD^i \Phi)^\dagger \Pi\bigr]\,. 
\ee
Using Eq.~\eqref{Pi} for $\Pi$, we obtain, 
\be 
\bP=-i\bigl[(\bD\Phi)^\dagger (\mu-eA_0) \Phi - \Phi^\dagger (\mu-eA_0)(\bD\Phi)\bigr]\,.\label{P}
\ee 
The kinetic angular momentum carried by the matter part is then 
\be 
L_z^{\rm matter}=\int_{\bx} \bigl[ \bx\times \bm P\bigr]_z \,. 
\label{eq:Lmatter}
\ee 
On the other hand, the gauge field contribution to the angular 
momentum is easily found from the Poynting vector, namely from Eq.~\eqref{eq:LEM}. 

The charge neutral case at $\mu=0$ makes the essential difference from
the nonrelativistic case in the previous section.  The above equations of motion for
relativistic vortices are mathematically identical, up to trivial
scaling of parameters, to those of ``locally'' neutral nonrelativistic
vortices without coupling to electric field:  both of them are known
as Nielsen-Olesen vortices.
For the charge neutral case, it is \textit{algebraically} trivial to
see $\Pi=0$ and $\bm P=\bm E=\boldsymbol{0}$ from $\mu=A_0=0$, and both the matter and gauge field 
contributions to the total angular momentum are zero, but this
conclusion is \textit{physically} nontrivial;  vortices are
circulating configurations and yet they have no angular momentum.
Intuitively, the absence of matter contribution to the angular
momentum in the relativistic theory can be understood as a
cancellation between particle vortex and anti-particle anti-vortex as
follows: recall that $\Phi\sim a+b^\dagger$ and
$\Pi\sim i(a-b^\dagger)$ where $a$ and $b$ are annihilation operators
for particle and anti-particle, respectively.  In the superfluid phase
of a large occupation number, we can regard $a$ and $b$ as c-numbers
as usual.  A vortex profile
with winding number $\nu$ can be viewed as a superposition of a 
particle vortex of $a\sim e^{i\nu\varphi}$ and an antiparticle 
antivortex of 
$b\sim e^{-i\nu\varphi}$.  In the charge neutral case of $\Pi=0$, 
their amplitudes are precisely equal, i.e., $a=b^\dagger$, and the 
antiparticle antivortex contribution to the angular momentum is 
precisely opposite to that of the particle vortex.  The absence of 
antiparticles in the nonrelativistic vortex in the previous section is 
the major difference from the relativistic theory discussed in this section. 

\begin{figure}
  \includegraphics[width=0.4\textwidth]{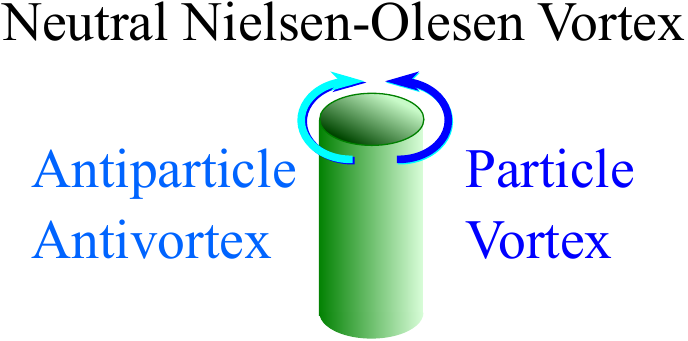}
  \caption{Schematic illustration of the charge neutral Nielsen-Olesen vortex 
    composed from a particle vortex and an antiparticle antivortex.}
  \label{fig:NO}
\end{figure}

For the charged case at $\mu\neq 0$, the two systems of equations are
different by terms that we previously neglected in the nonrelativistic
reduction.
In addition to this difference for the charged case, a background
charge density that we introduced as $q$ in the previous section is
also absent here.  This means that the net charge of a charged
Nielsen-Olesen vortices is not zero, and the electric field grows
logarithmically at large distance in two-dimensional space
perpendicular to the vortex string in three dimensions.  This implies
that the line density of energy of a charged vortex is divergent in
infinite space, and a sensible solution would exist only in a finite
transverse volume.  As we will focus on azimuthally symmetric vortex
configurations to apply our angular momentum conservation argument, we
consider a spatial cutoff in transverse space at a certain distance
from the origin, that is, $r\le R$ in the radial direction.  We will
show that the total angular momentum of a charged vortex within the
volume $r\le R$ is always zero for any cutoff $R$, when we sum the
contributions of both matter part and the gauge fields.

Our conclusion rectifies a misleading statement in the Appendix C of
the well-known literature, Ref.~\cite{Julia:1975ff}, that a charged
Nielsen-Olesen vortex carries a nonzero angular momentum.  Our result
of vanishing angular momentum even for charged vortices is independent
of the issue of diverging line energy density in infinite space.
Later, we will more precisely point out where the misleading
conclusion in Ref.~\cite{Julia:1975ff} stems from.

The computation in the charged vortex case is more delicate than the
neutral case, and the detailed mechanism for cancellation is similar
to that in the previous section.  First of all, since $\Pi\neq 0$, the
particle vortex and the antiparticle antivortex have different
amplitudes, and the net matter angular momentum no longer cancels to
be zero.  From the radial electric field, $\bE\neq \boldsymbol{0}$, the gauge
fields also contribute to the total angular momentum.  We take the
following Ansatz,
\be
  \Phi=f(r)\,e^{i\nu\varphi}\,,\qquad
  A_0=a(r)\,,\qquad
  A_{\varphi}={\nu\over e}\,\bigl[1-h(r)\bigr]
\ee
with the boundary condition for vanishing magnetic flux, i.e.,
$h(R)=0$ at sufficiently large boundary $r=R$.  Then the equations of
motion and the Gauss law become (with $'\equiv {d\over dr}$)
\begin{align}
  & {1\over r}(rf')' - {\nu^2\over r^2}h^2 f
    +(\lambda_2-\lambda_4 f^2)f+(ea-\mu)^2 f=0\,,\\
  & \biggl( {h'\over r}\biggr)' - {2e^2\over r}f^2 h=0\,,\\
  &  {1\over r}(ra')'-2e(ea-\mu)f^2=0\,.
    \label{Gausslaw}
\end{align}
The matter part of the angular momentum is computed from
Eq.~\eqref{eq:Lmatter} as
\be
  L_z^{\rm matter} = 4\pi\nu \int_0^R dr \, rh(r)f^2(r) \bigl[-ea(r)+\mu\bigr]
\ee
and the gauge field contribution from Eq.~\eqref{eq:LEM} as
\be
 L_z^{\rm gauge}=-{2\pi\nu\over e}\int_0^R dr\, ra'(r)h'(r)
={2\pi\nu\over e}\int_0^R dr\, [ra'(r)]' h(r)\,,
\ee
where in the last equality we performed the integration by part and
used the boundary condition at $r=R$ as in the previous section.
Using the Gauss law~\eqref{Gausslaw} to replace $[ra'(r)]'$, we arrive at
\be
L_z^{\rm gauge} = -4\pi\nu\int_0^R dr \, rh(r)f^2(r)[-ea(r)+\mu]
=-L_z^{\rm matter}\,,
\ee
which precisely cancels the matter contribution.  As a result the total
angular momentum is vanishing.  In Appendix~C of
Ref.~\cite{Julia:1975ff}, the surface term of Eq.~(C3) that was
neglected is nonzero:  this can be seen from the description of the
solution below Eq.~(C6) with $Q\neq 0$.  It can be shown that Eq.~(C3)
precisely cancels Eq.~(C4), so that the total angular momentum is
zero.  This cancellation has its origin in the angular momentum
conservation, and it holds for any $R$ regardless of an issue of infinite line
energy density of the charged solution.

We finish this subsection by pointing out that our finding of zero angular momentum for magnetic vortices in the $U(1)$ Abelian Higgs model is consistent with the particle-vortex duality in 2+1 dimensions \cite{Seiberg:2016gmd,Senthil:2018cru}, where the magnetic vortices in the Abelian Higgs model are mapped to the elementary excitations of a dual complex scalar field which are clearly spinless. Checking the spins of other excitations in the web of dualities \cite{Seiberg:2016gmd} would be interesting.

\subsection{Example 2: Non-Abelian CFL vortices}

\begin{figure}
  \includegraphics[width=0.35\textwidth]{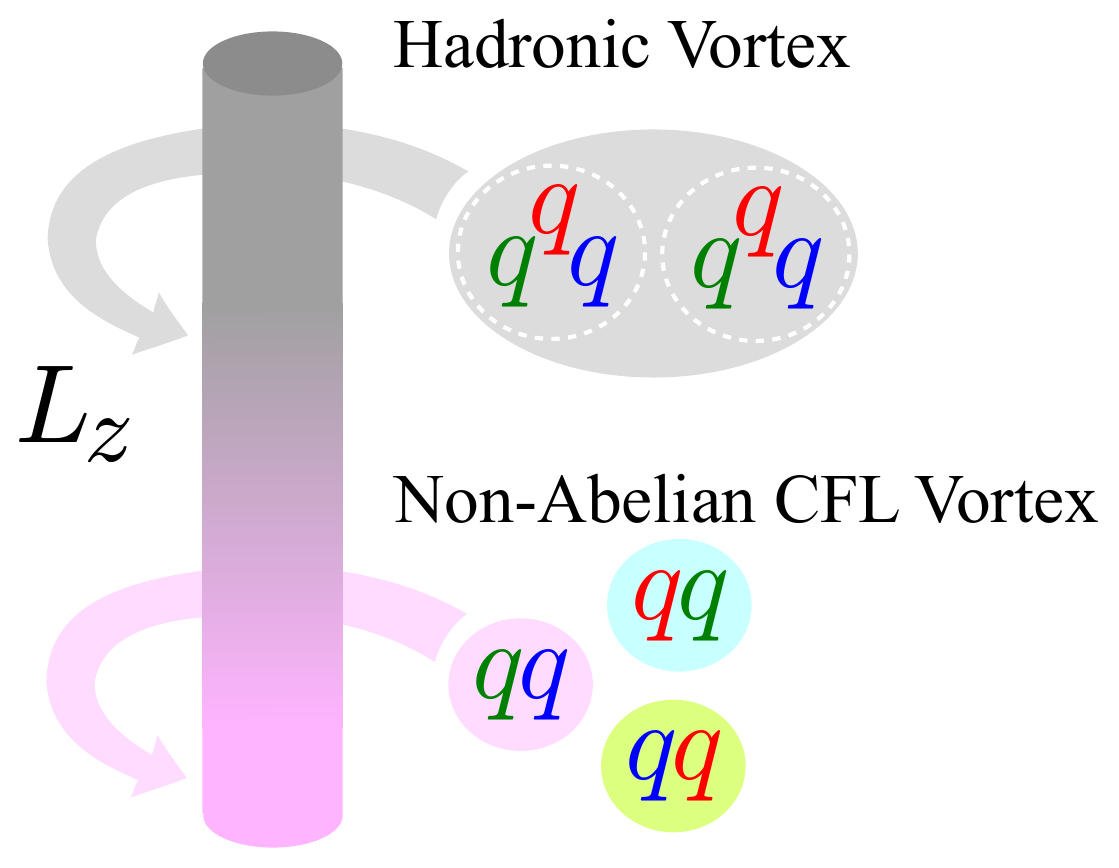}
  \caption{Schematic illustration of the continuity between the
    dibaryon vortex in the hadronic phase and the non-Abelian CFL
    vortex in CFL quark matter in QCD.}
  \label{fig:CFLvortex}
\end{figure}

We can test our assertion in a more non-trivial example of non-Abelian
vortices~\cite{PhysRevD.73.074009,PhysRevD.78.045002} in the CFL
color-superconducting phase of QCD quark matter at high baryon density
and low temperature.  The diquark condensates in the CFL phase break
both QCD gauge symmetry and the global $U(1)_{\rm B}$ baryon number
symmetry.  The non-Abelian vortices arise from coupled dynamics of
color fields and $U(1)_{\rm B}$ superfluidity, and carry fractional
winding numbers for both gauge and global symmetries, such that the
total winding number for each color component of the diquark
condensate field is an integer.  One might think that the non-Abelian
CFL vortex is peculiar to QCD, but similar structures can also be found in
multi-component superconductivity, see Ref.~\cite{Babaev:2001hv} for
example.  The minimal non-Abelian vortex carries only $1/2$ of the
$U(1)_{\rm B}$ winding number (that is equivalent to
${1\over 2}\times{2\over 3}={1\over 3}$ winding number for the diquark
field), so that the non-Abelian CFL vortices can be considered as
fractionalized $U(1)_{\rm B}$ vortices.  In the hadronic phase, on the
other hand, the minimal dibaryon Cooper-pair superfluid vortex also
carries the same winding number $1\over 2$, so that across the two
phases the dibaryon vortex should transmute to the non-Abelian CFL
vortex~\cite{Alford:2018mqj}, which is schematically illustrated in
Fig.~\ref{fig:CFLvortex} (see also Ref.~\cite{Chatterjee:2018nxe} for
an alternative scenario).  Since the angular momentum must be conserved
during this transmutation process, we expect the angular momenta of
the two vortices to be equal.  The minimal dibaryon vortex
of $1/2$ of the $U(1)_{\rm B}$ winding number is a usual superfluid
vortex and carries the angular momentum $L_z=N_{\rm B}/2$ where
$N_{\rm B}$ is the total baryon number.  In contrast, the non-Abelian
CFL vortex is also accompanied by color gauge fields, and in general,
its total angular momentum receives contribution from these localized
color fields.  It is a nontrivial check to see that the total angular
momentum of the non-Abelian CFL vortex from both matter part and the
gauge fields is indeed $L_z=N_{\rm B}/2$, i.e., the same as in the
hadronic phase, as we will show below.  Essentially,
this means that the color-magnetic part of the non-Abelian CFL vortex
does not contribute to the angular momentum, and only the
$U(1)_{\rm B}$ superfluid part makes a finite contribution.  This
situation provides another example of confirming our assertion that
the gauged magnetic vortex does not carry angular momentum.

The diquark condensate in the CFL phase is described by a $3\times 3$
matrix field, $\Phi=\Phi^{i\alpha}$, where $i$ and $\alpha$ are color
and flavor indices, respectively.  More precisely, there are two such
fields for left-handed and right-handed diquarks, and we assume that
they share the same configuration in a vortex solution.  We can always
perform suitable color rotations, such that the profile of the
non-Abelian CFL vortex appears only in the global $U(1)_{\rm B}$ and
the eighth component of the color field $A_\mu^8$ with the generator
$t^8={1\over \sqrt{12}}\mathrm{diag}(-2,1,1)$\footnote{This matrix
  representation is an unconventional choice;  in later discussions we
  will focus on the $u$-quark sector and this choice is good for
  that purpose.}.
Therefore, we will show expressions only in these parts in the
following.  The QCD covariant derivative with $A_\mu^8$ only is
\be 
D_\mu \Phi=\left(\partial_\mu -ig A_\mu^8 t^8\right) \Phi\,,
\ee 
where $g$ is the QCD coupling constant.  We will work with the gauge
invariant Lagrangian in terms of the diquark field given by
\be 
\calL={\rm tr}\bigl[ (D_0\Phi)^\dagger (D_0\Phi)
-(\bD\Phi)^\dagger (\bD\Phi)\bigr]
-V\left({\rm tr}(\Phi^\dagger\Phi)\right)
+{1\over 2}\bE^8\cdot\bE^8-{1\over 2}\bB^8\cdot\bB^8\,, 
\ee 
where $\bE^8=-\bnabla A_0^8-\partial_0 \bA^8$ and
$\bB^8=\bnabla\times\bA^8$.  The concrete shape of the potential $V$
is not important for our purpose.  The ensuing analysis is very
similar to that in the previous subsection, and we will highlight only
the important differences and the major results.

The chemical potential $\muB$ is introduced for the baryon charge
density $Q_{\rm B}$, which is 
\be 
Q_{\rm B} = -{2\over 3} i\, {\rm tr}
\bigl[ (D_0\Phi)^\dagger \Phi-\Phi^\dagger (D_0\Phi)\bigr]\,.
\ee 
Here the coefficient is understood from the baryon charge $2/3$
carried by the diquark field $\Phi$.  The color charge that appears in
the Gauss law constraint is given by 
\be 
\bnabla\cdot\bE^8 = ig\,{\rm tr}
\bigl[ (D_0\Phi)^\dagger t^8\Phi-\Phi^\dagger t^8 (D_0\Phi)\bigr]\,,
\ee 
which is easily obtained from the equation of motion for $A_0^8$. 
Introducing the canonical conjugate field $\Pi\equiv D_0\Phi$ and
following the same steps in the previous section, we can find the
Hamiltonian and the free energy as
\begin{align}
  F &={\rm tr} \bigl[\Pi^\dagger\Pi + (\bD\Phi)^\dagger(\bD\Phi)\bigr]
      +V\left({\rm tr}(\Phi^\dagger \Phi)\right) \notag\\
    &\quad + {g^2\over 2}{\rm tr}(\Pi^\dagger t^8\Phi-\Phi^\dagger t^8\Pi){1\over\bnabla^2}{\rm tr}(\Pi^\dagger t^8\Phi-\Phi^\dagger t^8 \Pi) \notag\\
    &\quad + {2\over 3}i \muB {\rm tr}(\Pi^\dagger\Phi-\Phi^\dagger\Pi )
      +{1\over 2}\bB^8\cdot\bB^8\,. 
\end{align}
Recall that the second line arises from the Coulomb energy of color
field, ${1\over 2}\bE^8\cdot\bE^8$ [see Eq.~\eqref{gauss}].  By
minimizing $F$ with respect to $\Pi$, $\Phi$, and $\bA^8$, we obtain
the equations of motion.  It is convenient to introduce an auxiliary
variable $A_0^8$ to render the nonlocality into local equations of
motion, which is defined by $\bE^8=-\bnabla A_0^8$ for static
configuration, so that the Gauss law becomes
\be 
\bnabla^2 A_0^8 = -ig{\rm tr}\left(\Pi^\dagger t^8 \Phi-\Phi^\dagger t^8 \Pi\right)\,.
\ee 
Then, the equation of motion for $\Pi$ is easily solved as
\be 
\Pi=-i(gA_0^8 t^8 +2 \muB/3)\Phi 
\ee 
and the Gauss law becomes 
\be 
\bm\nabla^2 A_0^8=2g\,{\rm tr}\bigl[\Phi^\dagger t^8(g A_0^8 t^8+2\mu_B/3)\Phi\bigr]\,.\label{CFL1}
\ee
In the above the term $\propto\muB$ should be understood as the unity
matrix in color space.  The other equations of motion are 
\be 
\bD^2 \Phi-\Phi V'\left({\rm tr}(\Phi^\dagger\Phi)\right)
+\bigl(gA_0^8 t^8 +2\muB/3\bigr)^2\Phi=0\,,\label{CFL2}
\ee 
and 
\be 
\bnabla\times(\bnabla\times \bA^8)=ig{\rm tr}
\bigl[(\bD \Phi)^\dagger t^8\Phi-\Phi^\dagger t^8 (\bD\Phi)\bigr]\,.
\label{CFL3}
\ee 
Equations~\eqref{CFL1}, \eqref{CFL2}, and \eqref{CFL3} form a closed set to
solve for the vortex profile of $\Phi$, $\bA^8$, and $A_0^8$. 
The non-Abelian CFL vortex solution has the following form~\cite{PhysRevD.73.074009}:
\be 
A_0^8=a(r)\,,\qquad
A^8_\varphi = {\nu\over g}{\sqrt{12}\over 3} \bigl[1-h(r)\bigr]
\ee
with $\bA^8= A^8_\varphi \hat{\varphi}/r$ and
\be 
\Phi=\begin{pmatrix}
  f(r)\,e^{i\nu\varphi} & 0 & 0 \\
  0 & b(r) & 0 \\
  0 & 0 & b(r)
\end{pmatrix}\,.
\label{eq:nonAbelian_f}
\ee 
The boundary condition is $h(\infty)=0$ which ensures,
\be 
-\bD\Phi=(\bnabla+ig\bA^8 t^8)\Phi ~~\to~~
i{\nu\over 3}\,{\hat{\varphi}\over r}\Phi
\quad \text{as } r\to\infty\,.
\ee
This signifies that the vortex carries a superfluid winding number
$\nu/3$ with respect to the diquark global $U(1)$ (which is equivalent to $\nu/2$ with respect to $U(1)_{\rm B}$ symmetry).  To see how the color-magnetic
vortex is embedded in the above solution, we can factorize it as follows,
\be
\Phi = e^{i{\nu\over 3}\varphi}
\begin{pmatrix} e^{i{2\nu \over 3}\varphi} & 0 & 0 \\
  0 & e^{-i{\nu\over 3}\varphi} & 0 \\
  0 & 0 & e^{-i{\nu\over 3}\varphi}
\end{pmatrix}
\begin{pmatrix} f(r) & 0 & 0 \\
  0 & b(r) & 0 \\
  0 & 0 & b(r)
\end{pmatrix}\,,
\ee 
where the overall phase corresponds to the global $U(1)$ and the middle
matrix, $e^{-i\nu{\sqrt{12}\over 3}t^8\varphi}$, belongs to $SU(3)$,
and this is why this configuration as implemented in
Eq.~\eqref{eq:nonAbelian_f} is called a ``non-Abelian'' vortex.

The equations of motion,  \eqref{CFL1}, \eqref{CFL2}, and
\eqref{CFL3}, become, after some algebra,
\bear 
&&{1\over r}(ra')'-{g^2\over 3} a(2f^2+b^2)-{4g\over 3\sqrt{3}}\mu_B(-f^2+b^2)=0\,,
\label{eq:Gauss_nonAbelian}\\
&&{1\over r}(rf')'-{\nu^2\over 9r^2}(1+2h)^2 f-f V'(f^2+2b^2)+\left(-{2g\over\sqrt{12}}a +{2\over 3}\mu_B\right)^2 f=0\,,\\
&&{1\over r}(rb')'-{\nu^2\over 9r^2}(1-h)^2 b-b V'(f^2+2b^2)+\left({g\over\sqrt{12}}a +{2\over 3}\mu_B\right)^2 b=0\,,\\
&& \left({1\over r}h'\right)'+{g^2\over 3}{f^2\over r}(1+2h) -{g^2\over 3r}(1-h)b^2 =0\,,
\eear 
where Eq.~\eqref{eq:Gauss_nonAbelian} corresponds to the Gauss law.

To compute the matter part of the angular momentum, we need the 
momentum density, 
\be 
\bP^i=T^{0i}={\rm tr}\bigl[(D_0 \Phi)^\dagger (\bD^i \Phi) 
+(\bD^i\Phi)^\dagger (D_0\Phi)\bigr]={\rm tr}\bigl[\Pi^\dagger 
(\bD^i \Phi)+(\bD^i \Phi)^\dagger \Pi\bigr]\,. 
\ee 
Substituting the solution of $\Pi$ for the above $\bP^i$, we obtain, 
\begin{align}
  \bP &= P_\varphi\hat{\varphi}
        = -i\,{\rm tr}\bigl[(\bD\Phi)^\dagger (gA_0^8 t^8 +2\muB/3) \Phi 
        - \Phi^\dagger (gA_0^8 t^8 +2\muB/3)(\bD\Phi)\bigr] \notag\\
      &=\biggl[ {\nu\over 3r}(1+2h){f^2} \biggl(-{4g\over\sqrt{12}} a
        +{4\over 3}\muB\biggr) + {4\nu\over 3r}(1-h)b^2
        \biggl( {g\over\sqrt{12}}a+{2\over 3}\muB\biggr)
        \biggr]\hat\varphi\,.
\end{align}
Thus, the matter part of the angular momentum is 
\begin{align}
  L_z^{\rm matter}
  &=2\pi\int_0^R dr\,\, r (rP_\varphi) \notag\\
  &= {2\pi\nu \over 3}\int_0^R dr\,
    r\biggl[ (1+2h){f^2}\biggl(-{4g\over\sqrt{12}} a+{4\over 3}\muB\biggr)
    +{4}(1-h)b^2\biggl( {g\over\sqrt{12}}a+{2\over 3}\muB\biggr)
    \biggr]\,. \label{Lzm}
\end{align}
The color gauge field contribution is as before
\be 
L_z^{\rm gauge}=-(2\pi){\sqrt{12}\nu\over 3g}\int_0^R dr\, r a'(r)h'(r)\,.
\ee 
Integrating by part and using the Gauss law, we have, 
\be 
L_z^{\rm gauge}=(2\pi\nu){\sqrt{12}\over 3}\int_0^Rdr\,
r\biggl[ {g\over 3} a(2f^2+b^2)+{4\over 3\sqrt{3}}\muB (-f^2+b^2)\biggr]h\,. \label{Lzg}
\ee 
Summing up $L_z^{\rm matter}$ in Eq.~\eqref{Lzm} and $L_z^{\rm gauge}$ in
Eq.~\eqref{Lzg}, we get the total angular momentum per unit vortex
length to be
\be 
L_z^{\rm tot} = 2\pi\nu\int_0^R dr\,r\left[f^2\left(-{2g\over 3\sqrt{3}}a+{4\over 9}\muB\right)+b^2\left({2g\over 3\sqrt{3}}a+{8\over 9}\muB\right)\right]\,.
\ee 
One might think that the above result is an involved expression, but
there is an elegant physical interpretation.  To this end, we shall
compute the baryon charge density as
\bear 
Q_{\rm B} &=& -{2\over 3}i\,{\rm tr}
\bigl[ (D_0\Phi)^\dagger \Phi-\Phi^\dagger (D_0\Phi)\bigr]
=-{2\over 3}i\,{\rm tr} (\Pi^\dagger \Phi-\Phi^\dagger\Pi )\nonumber\\
&=& {4\over 3}{\rm tr}\bigl[ \Phi^\dagger (gA_0^8 t^8+2\muB/3)\Phi\bigr]\,,
\eear 
from which the total baryon charge per unit vortex length reads:
\be 
N_{\rm B}=2\pi\int_0^R dr\,r\left[f^2\left(-{4g\over 3\sqrt{3}}a+{8\over 9}\muB\right)+b^2\left({4g\over 3\sqrt{3}}a+{16\over 9}\muB\right)\right]\,.
\ee 
Comparing $L_z^{\rm tot}$ and $N_{\rm B}$, we see that the following
relation holds:
\be 
L_z^{\rm tot} = {\nu\over 2}N_{\rm B}\,.
\ee 
This confirms that the total angular momentum of the non-Abelian
vortex in the CFL phase contains only the contribution from the global
$U(1)_{\rm B}$ vortex;  the total angular momentum (in the unit of
$\hbar$) is $\nu$ times the number of the Cooper pairs.  We note that
this result completely agrees with Eq.~(14) in
Ref.~\cite{Alford:2018mqj}, but in Ref.~\cite{Alford:2018mqj} only the
$U(1)_{\rm B}$ contribution to the angular momentum was postulated
without rigorous justification.

\section{Class III: Case study without angular momentum conservation}

The last logical possibility in our classification is that magnetic
vortices cannot be created by simply turning on external magnetic flux
in an azimuthally symmetric way.  What distinguishes this case from
all previous cases is that the principle of angular momentum
conservation does not na\"{i}vely apply in the vortex creation
process.  The vortices classified in this class are characterized by
inhomogeneous profiles along the vortex axis, which means that not
only the external magnetic flux but also something else are needed to
create the vortices:  roughly speaking, a kind of twisting along the
axis would be required.  Such vortices do exist as we discuss below,
although they seem to be rare in the literature.

An example that belongs to this class is provided by an object called
``charged semilocal vortex'' as constructed by Abraham in
Ref.~\cite{Abraham:1992hv}.   This Abraham vortex is an extension of
the semilocal vortex~\cite{Vachaspati:1991dz}  that has been discussed
in the context of electroweak strings in cosmology (see
Ref.~\cite{Achucarro:1999it} for a review).  They also appear quite
commonly as topological BPS (Bogomol'nyi-Prasad-Sommerfield) objects
in supersymmetric gauge theories.  The simplest model of the
Abraham vortex consists of two charged scalar fields, $\Phi_{a}$ $(a=1,2)$,
with the equal charge, and a $U(1)$ gauge field $A_\mu$.  The
Hamiltonian in the critical limit reads:
\be
H=\sum_{a=1,2}\left(|D_0\Phi_{a}|^2+|\bD\Phi_{a}|^2\right)+{g^2\over 2}
\biggl(\,\, \sum_{a=1,2}|\Phi_{a}|^2-v^2\biggr)^2+{1\over 2}(\bE^2+\bB^2)\,,
\ee
where $D_\mu\Phi_a=(\partial_\mu-igA_\mu)\Phi_a$, and the Gauss law is
\be
\bnabla\cdot\bE=ig \sum_{a=1,2}\Bigl[ (D_0\Phi_a)^\dagger \Phi_a-\Phi_a^\dagger (D_0\Phi_a)\Bigr]\,.
\ee
The vortex solution relies on the following Bogomol'nyi bound,
\begin{align}
H&=\sum_{a=1,2}\bigl(|D_0\Phi_{a}\pm D_3\Phi_{a}|^2+|D_1\Phi_{a}\pm i D_{2}\Phi_{a}|^2\bigr)+{1\over 2}|E_x\mp B_y|^2+{1\over 2}|E_y\pm B_x|^2\notag\\
&\quad +{1\over 2}\biggl[ B_z\mp g\biggl(\sum_{a=1,2}|\Phi_a|^2-v^2\biggr)\biggr]^2\mp \alpha Q_2\mp v^2 g B_z\mp \bm\nabla\cdot(\bm E A_z)\,.
\end{align}
Here, $\bA=(A_x,A_y,A_z)$, $\bE=(E_x,E_y,E_z)$, $\bB=(B_x,B_y,B_z)$ and 
we imposed additional conditions that
$\partial_3\Phi_2=i\alpha \Phi_2$ with a constant $\alpha$ and all
other $\partial_3$ is vanishing.  We defined $Q_2$ as
\be
Q_2=i\bigl[ (D_0\Phi_2)^\dagger \Phi_2 - \Phi_2^\dagger (D_0\Phi_2)\bigr]\,.
\ee
The equations we obtain from this, for the upper sign, are
\be
D_0\Phi_{a}+ D_3\Phi_{a}=0\,,\quad D_1\Phi_{a}+ i D_{2}\Phi_{a}=0\,,\quad E_i=\epsilon_{ij}B_j\,,\quad B_z- g\biggl(\sum_{a=1,2}|\Phi_a|^2-v^2\biggr)=0\,.
\ee
It can be checked that these solve the original equations of motion.
In Ref.~\cite{Abraham:1992hv} it was also shown that these equations
admit nice solutions with zero net gauged $U(1)$ charge but nonzero
$Q_2$, which are somewhat misleadingly called ``charged'' semilocal
vortices.  These solutions have finite line energy density, due to the
fact that the total $U(1)$ charge is zero.  The solutions are possible
only for $\alpha\neq 0$, that corresponds to a ``twisting''
along the vortex axis.  Due to this, the vortex string carries a net
linear momentum along the axis direction.  Although the total $U(1)$
charge is zero, the charge density profile in space is nonzero, and
there exists nontrivial profile of local electric and magnetic fields.
This leads to a nonvanishing contribution of the electromagnetic
fields to the total angular momentum.  As pointed out in
Ref.~\cite{Abraham:1992hv}, the solutions carry nonzero total angular
momentum, but we would not go into technicality here, and the readers
can directly consult Ref.~\cite{Abraham:1992hv}.  The ``twisting'',
parametrized by $\alpha$, can be considered as spinning the vortex to
give a finite angular momentum.  This is an extra operation that would
be needed to create such a vortex profile by hand, and the angular
momentum conservation cannot be applied to the situation.  In other
words, in this peculiar system belonging to this class, the angular
momentum conservation is not satisfied by $L_z^{\rm matter}$ or
$L_z^{\rm gauge}$ or their sum. 

\section{Conclusion}

In this work, we apply the principle of angular momentum conservation
to understand the origin of angular momentum carried by magnetic
vortices in various physical systems in condensed matter, high energy,
nuclear physics, and cosmology.  We find that this simple principle is
powerful enough to allow us an overarching scheme of classifying the
known examples, according to how the principle of angular momentum
conservation is satisfied.  We find the four distinct classes of
examples in our classification scheme; spinful (class Ia), topological
(class Ib), spinless (class II) and exotic (class III) vortices, as
already summarized in Introduction.  We present detailed analyses for
these examples, and emphasize that the angular momentum carried by
localized gauge fields around the vortex core plays a crucial role in
satisfying the angular momentum conservation.  We believe that our
study gives a clear answer to the seemingly confusing, but
surprisingly rich, question of angular momentum carried by magnetic
vortices that are ubiquitous in many branches of physics.

\acknowledgements

K.~F.\ is supported by Japan Society for the Promotion of Science
(JSPS) KAKENHI Grant Nos.\ 18H01211 and 19K21874.
Y.~H.\ is supported by Japan Society for the Promotion of Science
(JSPS) KAKENHI Grant Nos.\ 17H06462 and 18H01211.
H.-U.~Y.\ is supported by the U.S. Department of Energy, Office of
Science, Office of Nuclear Physics, Grant No. DE-FG0201ER41195, and
within the framework of the Beam Energy Scan Theory (BEST) Topical
Collaboration.

%
%
%
%
%
%
%
%
%
%
%
%
%
%
%

\bibliography{vortexAM}

\end{document}